\documentclass[manuscript,screen,natbib=false]{acmart}
\usepackage[utf8]{inputenc}
\usepackage{graphicx}
\usepackage{algorithm}
\usepackage{algpseudocode}
\usepackage{booktabs}
\usepackage{xcolor}
\usepackage[inline]{enumitem}
\usepackage{makecell}

\makeatletter
  \def\@formatdoi#1{}%
  \def\@acmVolume{\hspace{-.6ex}}
\makeatother
\acmJournal{TAAS}
\acmNumber{Submitted}
\setcopyright{cc}
\copyrightyear{2026}
\acmYear{2026}

\usepackage[abbreviate=true, dateabbrev=true, alldates=year, isbn=false, doi=true, urldate=comp, url=false, maxbibnames=3, backref=false, backend=biber, style=numeric-comp, language=american, giveninits=true, sortcites=true, sorting=none, natbib=false]{biblatex}
\DeclareSourcemap{
  \maps[datatype=bibtex]{
    \map{
      \step[fieldsource=booktitle, match={\{\{?}, replace={}] 			
      \step[fieldsource=booktitle, match={\}\}?}, replace={}] 	
      \step[fieldsource=booktitle, match=\regexp{^Proceedings\s+of\s+the\s+}, replace={}]
      \step[fieldsource=booktitle, match=\regexp{^\d\d\d\d\s+}, replace={}]
      \step[fieldsource=booktitle, match=\regexp{^[\d]{1,2}th\s+}, replace={}]
      \step[fieldsource=booktitle, match=\regexp{^[\d]{1,2}nd\s+}, replace={}]
      \step[fieldsource=booktitle, match=\regexp{^SIG[A-Z]+\s+}, replace={}]
      \step[fieldsource=booktitle, match={International}, replace={Int'l}]
      \step[fieldsource=booktitle, match=\regexp{Conference(\s+on)?(\s+the)?}, replace={Conf.}]
      \step[fieldsource=booktitle, match=\regexp{Symposium(\s+on)?(\s+the)?}, replace={Symp.}]
      \step[fieldsource=booktitle, match=\regexp{Workshop[s]?(\s+on?)(\s+the)?}, replace={Ws.}]
      \step[fieldsource=booktitle, match=\regexp{\.\s+}, replace=\regexp{.\\\x{20}}]
      \step[fieldsource=journal, match=\regexp{(The\s+)?Journal\s+of\s+}, replace=\regexp{J.\\\x{20}}]
      \step[fieldsource=journal, match=\regexp{Communications\s+of\s+the\s+}, replace=\regexp{Comm.\\\x{20}}]
      \step[fieldsource=journal, match=\regexp{(ACM|IEEE)\s+Transactions\s+on\s+(.*)}, replace=\regexp{$1\x{20}Trans.\\\x{20}$2}]
      \step[fieldsource=journal, match={Advances in}, replace={Adv.\ }]
      \step[fieldsource=journal, match=\regexp{\.\s+}, replace=\regexp{.\\\x{20}}]
      \step[fieldsource=publisher, match={Association for Computing Machinery}, replace={ACM}]
      \step[fieldsource=publisher, match={Springer-Verlag}, replace={Springer}]
      \step[fieldsource=publisher, match=\regexp{(ACM|IEEE)\s+Press}, replace=regexp{$1}]
      \step[typesource=techreport, typetarget=report]
    }
  }
}

\newcommand{\RemoveArxivDOIs}{
  \DeclareSourcemap{
    \maps[datatype=bibtex]{
      \map{
        \pertype{misc}
        \step[fieldsource=publisher, match=\regexp{(?i)^arxiv$}, final]
        \step[fieldset=doi, null]
        \step[fieldset=primaryclass, null]
      }
    }
  }
}

\newcommand{\RemoveBibField}[2]{
  \DeclareSourcemap{\maps[datatype=bibtex]{\map{\step[fieldsource=entrykey, match={#1}, final]\step[fieldset=#2, null]
}}}}

 \AtEveryBibitem{\clearlist{location}} %
\AtEveryBibitem{\clearfield{series}} %
\AtEveryBibitem{\clearlist{language}} %
\AtEveryBibitem{\clearfield{note}} %
\RemoveArxivDOIs{} %
\RemoveBibField{achiam2017:constrained}{publisher}

\newcommand{\tablefontsize}{\footnotesize}
\newcommand{\method}{CRRL}

\newcommand{\head}[1]{\par\noindent\textbf{#1:}\space}

\newcommand{\SimulaAffiliation}{%
  \affiliation{%
    \institution{Simula Research Laboratory}
    \city{Oslo}
    \country{Norway}
  }
}
\newcommand{\UiOAffiliation}{%
  \affiliation{%
    \institution{University of Oslo}
    \city{Oslo}
    \country{Norway}
  }
}

\begin{document}

\title{\method{}: A Causality-Based Reinforcement Learning Framework for Autonomous System Recovery}

\author{Safia Fatima}
\email{safia@simula.no}
\orcid{0000-0002-0706-677X}
\SimulaAffiliation{}
\author{Kai Olav Ellefsen}
\orcid{0000-0003-2466-2319}
\email{kaiolae@ifi.uio.no}
\UiOAffiliation{}
\author{Leon Moonen}
\orcid{0000-0002-1761-6771}
\email{leon.moonen@computer.org}
\SimulaAffiliation{}
\authornote{~Corresponding author.}

\begin{abstract}
Traditional reinforcement learning (RL) for recovery in autonomous systems lacks causal understanding and generalizes poorly to novel failure scenarios. RL policies often stall in failure states, spending up to 70\% of an episode immobilized. Rule-based recovery alone is inadequate, and adding heuristic recovery to a pretrained PPO policy \emph{worsens} rewards because policies cannot coordinate well with unanticipated interventions. The issue is not missing recovery mechanisms but a lack of policies trained to collaborate with them. 

We introduce CRRL, a causal-guided RL framework that trains policies to work effectively with rule-based recovery.
The recovery detects stalled states and assists the agent. Causal relations from driving logs shape the training signal, teaching the policy to anticipate stalls and adjust actions in recovery contexts. The framework follows MAPE-K, with sensor collection, causal model construction, and hybrid RL policy training corresponding to Monitor, Analyze, and Plan/Execute, respectively.

We evaluate CRRL through a four-condition ablation study across three driving scenarios, with 20 episodes per condition. We find that causal training significantly improves reward, distance, and velocity. Moreover, 9 of 20 roundabout episodes required zero recovery intervention, confirming navigation competence. These results show that causal-guided training produces effective RL policies that cooperate with rule-based safety components.

\end{abstract}



\maketitle

\section{Introduction}

Modern autonomous systems are increasingly deployed in complex, unpredictable environments where failures can have severe consequences. From autonomous vehicles navigating busy urban environments to financial trading systems processing billions of dollars in transactions, the ability to detect and recover from failures in real-time has become a critical requirement. This challenge is central to the field of self-adaptive systems, where a managing system monitors, analyses, and adapts a managed component at runtime to maintain acceptable behaviour~\cite{weyns2012:forms,psaier2011:survey}.
Because failures cannot be fully eliminated in complex real-world environments, therefore we focus on systems that cooperate effectively with recovery mechanisms when failures occur.

Traditional failure recovery typically relies on rule-based systems or reinforcement learning methods that learn appropriate recovery actions from extensive trajectory data~\cite{ahmadzadeh2014:multiobjective,akowuah2021:recoverybylearning}. 
While these approaches show promise in controlled environments, they face significant challenges when deployed in real-world scenarios, especially when dealing with rare or previously unseen failure events. Their core limitations are a lack of causal understanding of system dynamics and an inability to train policies that cooperate with external recovery mechanisms. A key finding of our work illustrates the latter that adding a heuristic reverse-recovery module to a pretrained PPO policy degrades reward across all evaluated scenarios because the policy cannot cooperate productively with interventions it was not trained to anticipate. %

Causal inference offers a principled framework to model cause-and-effect relations in complex systems, enabling reasoning about interventions and counterfactuals for robust decision-making under uncertainty. Prior work has combined causal reasoning with reinforcement learning, Causal Inference Q-Network (CIQ) handles observational noise in gaming~\cite{yang2021:causal}, and Gasse et al.\ reframe model-based RL via do-calculus in synthetic settings~\cite{gasse2021:causal}, but neither addresses physical failure recovery in autonomous driving. In driving contexts, causal discovery has been used for imitation learning~\cite{samsami2021:causal} and for counterfactual reasoning to understand agent behavior~\cite{howard2025:extending}, but neither addresses failure recovery. To the best of our knowledge, no existing work uses causal models from real driving interactions to enable learned policies to cooperate with rule-based recovery mechanisms. By integrating causal structure with RL during training, we can teach policies cooperative behaviors that rule-based recovery mechanisms alone cannot provide.

We address this gap by introducing Causal Recovery Reinforcement Learning (CRRL), a hybrid framework that combines causal-guided reinforcement learning with a rule-based recovery module. The causally trained PPO policy manages forward driving, obstacle avoidance, anticipation of stalled states, and post-recovery navigation, while the rule-based module handles reverse maneuvers using velocity-based heuristics, choosing actions via heuristic cycling or causal risk-ranked selection. Causal guidance shapes the policy parameters during training rather than correcting at runtime. 
The main contributions of this work are as follows:
\begin{itemize}[label=\raisebox{1pt}{\footnotesize$\bigstar$},nosep]

\item\textbf{\method{}}: 
a hybrid framework where causal models learned from driving logs shape the RL training signal through a multimodal reward design, encouraging a PPO policy to collaborate with a rule-based recovery module.

\item A \textbf{three-stage pipeline} that couples Bayesian-network causal modeling with PPO: trajectory data collection, causal model construction, and causally informed policy training, where causal guidance is applied at \textit{training time} to influence policy parameters, enabling cooperative behavior between the policy and recovery module.

\item We provide controlled \textbf{empirical evidence} across three increasingly complex driving scenarios: (i)~a \textit{straight road} with two opposing vehicles testing baseline recovery, (ii)~a \textit{roundabout} with curved sections and angled intersections testing steering under sustained curvature, and (iii)~a \textit{T-junction} with perpendicular approach roads testing recovery under sharp directional changes. 

\item We conduct a \textbf{four-condition ablation study} across the three scenarios to isolate the contribution of each component in our pipeline, and find that causal-guided policy training is the primary source of improvement.

\item We provide a \textbf{replication package} for our experimental framework, including raw data and analysis scripts.\footnote{~Privately shared at \url{https://figshare.com/s/40f7382ee937af853afd} while under review; it will be published on Zenodo after paper acceptance.}

\end{itemize}

\section{Background and Literature Review} \label{sec:RW}
As technology becomes more embedded in everyday life, the importance of software systems to recover autonomously increases. Prior work has investigated this problem from multiple angles. Machine learning and anomaly-detection approaches uncover faults and enable self-healing in networked and IoT  systems~\cite{zidi2024:fault,devi2024:selfhealing}. A broad range of self-adaptive solutions has been proposed, including rule-based, ML-based, and evolutionary approaches~\cite{salehie2009:selfadaptive,weyns2012:forms,psaier2011:survey}.

Reinforcement learning (RL) has been applied to recovery tasks across domains, from underwater vehicles to cyber-physical systems~\cite{ahmadzadeh2014:multiobjective,akowuah2021:recoverybylearning,thananjeyan2021:recovery}. 
In the self-adaptive systems (SAS) community, this is typically framed as a managed/managing system decomposition~\cite{weyns2012:forms}, in which a managing system monitors the managed system, analyzes its state, plans adaptations, and executes them. This decomposition is reflected in \method{}’s architecture. A key driver of self-adaptation is managing uncertainty during operation~\cite{hezavehi2021:uncertaintya}, where identifying uncertainty in monitoring and decision-making is recognized as a primary open problem, one that we aim to address in this work through causal probability estimation.

In this review, we focus on the recovery of software systems utilizing RL and causal inference. RL is usually formulated as a Markov Decision Process (MDP), a framework that enables agents to learn optimal behaviors in a given environment. The core objective is to maximize cumulative rewards through environmental interactions \cite{kaelbling1996:reinforcement}, and causal inference is the process of identifying and quantifying the causal effect of one variable on another. It involves using statistical methods, study designs, and theoretical frameworks to establish causality while accounting for confounding factors, potential biases, and the limitations of observational data~\cite{pearl2010:causal}. A central tool in causal inference is the \textit{do-operator}, introduced by Pearl~\cite{mcdonald2002:judea, peters2017:elements}. While conditional probability $P(Y \mid X = x)$ captures statistical association (including confounded relationships), the interventional distribution $P(Y \mid do(X = x))$ represents the effect of actively setting $X$ to $x$, removing all other influences on $X$. This distinction between observation and intervention is fundamental to causal reasoning, enabling the system to predict the consequences of its own actions rather than merely observing correlations. %
In the context of autonomous driving recovery, querying $P(\text{collision} \mid do(\text{action} = a))$ would estimate the causal effect of taking action $a$, rather than the mere correlation between that action and collisions in the training data. In \method{}, we approximate this reasoning by performing conditional inference over a causal DAG whose directed edges encode  cause--effect relationships, yielding $P(\text{collision} = 1 \mid \text{action},\, \text{state})$(Section~\ref{sec:causal_risk_ranked})

\subsection{Reinforcement Learning}
In recent years, RL has been explored for many areas, including signal control, microservices, recovery, fault tolerance, and many more.
Researchers have applied deep reinforcement learning (DRL) to traffic signal control using a novel action representation built from an inexperienced action set~\cite{zhancheng2021:researcha}. This approach presents a novel way of modeling deep reinforcement learning within the context of traffic signal control, however the state transmission and action space assume ideal conditions. There is a need for a more practical and well-considered training environment to validate the findings further and improve their applicability.
DeepDRAMA applies DRL to disaster recovery in elastic optical networks ~\cite{zou2021:deepdrama}. It defines a "mitigation zone" around the disaster area to enable controlled service degradation during traffic re-routing. Using Deep Q-learning, it optimizes degradation levels for affected demands, representing an early use of DRL for optical network disaster management. Simulations on COST239 and NSF topologies demonstrate strong performance, with faster gains in larger networks and training convergence in roughly 100 episodes. Future work aims to include joint optimization of path and spectrum reassignment.

Offline RL has been explored for automated error recovery by deriving recovery rules from existing policies~\cite{zhu2007:reinforcement}. Combining Q-learning with learned and user-specified policies reduced cluster downtime by 10\% and converged faster (~40,000 sweeps), but it depends on recovery log coverage and can yield locally optimal behavior.
A multi-agent fault-recovery architecture for self-healing systems separates detection, diagnosis, planning, and execution~\cite{rajput2021:multiagent} on JADE, it restored unavailability in 100--150 ms and handled over-allocation in 200--1200 ms, though evaluation is limited and single-threaded.
For microservices, MicroRAS avoids historical failure traces by using real-time monitoring to estimate action risk and benefit via a graph-based state model, action-effect estimator, and fuzzy selector~\cite{wu2020:microras}. It reports 94.7\% recovery, 44.3\% less interference, and 4x faster mitigation, making it useful when logs are sparse.

Recovery RL~\cite{thananjeyan2021:recovery} is a safe RL method that pairs a task policy with a recovery policy that intervenes near constraint violations. It uses offline data to pre-train safety components to reduce unsafe exploration. Results show 2--20 times higher sample efficiency in simulation and about 3 times gains on real robots, with open questions on broader validation and stronger guarantees.
More generally, safe RL studies how to learn effective policies while satisfying constraints during training and deployment~\cite{garcia:comprehensive}. This objective is closely related to \method{}, which incorporates estimated failure probabilities into the reward to bias learning away from high-risk behaviors.
In self-adaptive and self-healing systems, RL has been used to select repair actions online via utility-guided planning in large, dynamic architectures~\cite{ghahremani2020:improving}. Building on this line of work, we target physical autonomous vehicles and introduce causal structure into the learning signal, enabling tighter coordination with a concurrent rule-based recovery module. Related efforts reduce the size of adaptation spaces using deep learning to prioritize promising candidates for evaluation~\cite{weyns2022:deep}, which conceptually aligns with \method{}'s causal risk-based ranking of recovery actions.

Early DRL for autonomous driving used deep deterministic policy gradient (DDPG) in Open Racing Car Simulator with a 29D input and custom reward~\cite{wang2019:deep}. It showed promise but struggled with collision avoidance. Later work improves PPO in CARLA via multi-objective reward design~\cite{song2023:autonomous}. \method{} extends this with causal reward shaping, using a Bayesian network to estimate collision risk.
Surveys highlight sim-to-real transfer, sample efficiency, and safety as key challenges in DRL driving~\cite{kiran2022:deep}. The Chauffeurnet case study shows that even 30M expert samples can be insufficient. In mixed autonomy, RL policies must adapt to rule-following human drivers~\cite{valiente2022:robustness}. \method{} similarly requires PPO to coordinate with a rule-based recovery module.

Prior work shows that conventional RL for autonomous systems struggles with rare failures, weak causal understanding of action outcomes, and limited sim-to-real transfer, motivating our \emph{causal-guided hybrid design} of \method{}.

\subsection{Causal Inference}

Recent work targets out-of-distribution (OOD) conditions in autonomous-vehicle safety~\cite{filos2020:can}. RIP detects and responds to distribution shifts, AdaRIP adds real-time expert feedback, and CARNOVEL benchmarks novel conditions. The approach combines Bayesian imitation modeling with ensemble shift detection and evaluates both worst-case and model-average objectives, but remains constrained by incomplete shift coverage, high computational cost, and reliance on expert availability.
Causal RL frameworks integrate observational and interventional data to learn causal models in partially observable settings~\cite{gasse2021:causal}. They recast model-based RL as causal inference, using do-calculus to combine offline and online data via a joint latent-variable model that addresses confounding and offers theoretical guarantees, improving generalization and learning efficiency in experiments. Evidence is still largely synthetic, and modeling latent confounders in complex real-world systems remains open.

Causal reasoning can strengthen multi-agent RL by modeling agent interactions with explicit cause–and–effect structure using Structural Causal Models (SCMs) and Multi-Agent Causal Models (MACMs)~\cite{grimbly2021:causal}. Key challenges include non-stationarity, knowledge sharing, credit assignment, and confounding, motivating a “causality-first” approach based on Pearl’s ladder of causation. While promising for data efficiency, interpretability, and guarantees, the field remains largely theoretical. Open problems include tightly integrating graphical causal tools with MARL and learning reliable causal models in multi-agent settings.
Benchmarking work on temporal observational causal discovery for autonomous driving assesses inference of causal relations among road agents under sparse, non-stationary dynamics~\cite{howard2023:evaluating}. Methods perform well on synthetic convoy-style scenarios but degrade on real-world data, especially under sparsity and non-stationarity. Current benchmarks are limited in scenario complexity and dataset diversity, and they emphasize the need for approaches that transfer to realistic driving conditions.
Counterfactual simulation supports causal discovery among autonomous-driving agents by inferring cause–and–effect links in speed selection and benchmarking against observational methods~\cite{howard2023:simulationbased},. Agent-based and hybrid variants perform best, but the runtime is high (3.16 s) and the reward-based variant needs optimization.
Extensions to SCMs add mechanisms for modularization and fixed-size temporal representations, enabling collision fault attribution via counterfactual simulation~\cite{howard2025:extending}. The work is still largely theoretical, with open needs for validation and real-time integration.

TraffNet enables real-time what-if traffic prediction by learning causal structure in traffic flow from vehicle trajectories, using an OD-informed heterogeneous graph plus temporal modeling~\cite{xu2024:traffnet}. On Sumo-SY, it improves RMSE by 15.88\% and MAE by 18.01\%, and is robust under accidents, but currently supports only relatively simple interventions (an open need is extending to traffic signal control).
CityLifeSim is an Unreal Engine and AirSim urban simulation environment with configurable pedestrians and events. Evaluations show strong performance in standard views but degraded results in drone views and adverse weather, and it still needs stronger real-world validation~\cite{wang2022:citylifesim}.
Causal City is a high-fidelity AirSim and Unreal simulation for causal discovery and reasoning in autonomous driving, logging multimodal data. Benchmarking shows prediction error grows over time, especially in the more complex dataset, and realism and completeness remain limitations~\cite{mcduff2022:causalcity}.

CIM (Causal Imitative Model) improves autonomous driving by learning causal structure from expert demonstrations to reduce collisions and improve OOD performance\cite{samsami2021:causal}. It uses a disentangled latent space to extract interpretable causal variables and a causal pipeline for perception and control. In CARLA, it reports 69\% lower error than baselines, but remains limited by imperfect disentanglement and incomplete scene-navigation. 
IRCA identifies root causes of failures in autonomous-driving perception stacks by modeling module dependencies with a hierarchical SCM and using counterfactual interventions~\cite{wang2025:interventional}. It reports 95.39\% F1 on real faults and 96.73\% on synthetic faults in 1.58 minutes, but currently handles only OR-type fault relations, AND-type is refered as future work. 

The literature offers several methods for inferring causal directed acyclic graphs (DAGs) from observational logs. Constraint-based approaches, such as the PC algorithm~\cite{spirtes1993:causation}, use conditional independence tests to prune and orient edges, while score-based approaches, such as GES~\cite{chickering:optimal}, search over candidate graphs to maximize a structure score (BIC). Hybrid methods combine these ideas to balance robustness and search efficiency.
For time-series data, Granger causality~\cite{granger1969:investigating} tests whether lagged values of one variable improve the prediction of another, and has been used in driving-related settings~\cite{samsami2021:causal}. More recent methods, including PCMCI~\cite{runge2019:detecting}, extend these tests to better handle nonlinear relationships and high-dimensional sets of variables. In parallel, Bayesian structure learning, such as CausalNex~\cite{:welcome}, supports learning DAGs from discretized observational data while incorporating expert knowledge through structural constraints. We follow this latter direction, building our causal graph in CausalNex and imposing domain-specific edge constraints based on autonomous driving knowledge (detailed in Section~\ref{sec:causal_graph}). The resulting causal Bayesian network in \method{} also serves as a learned world model~\cite{moerland2023:modelbased}, which we use during training to estimate failure probabilities and shape the reward signal.

\head{Research Gap} RL shows promise for recovery systems, but current methods still struggle with rare failures, limited causal understanding, and dependable real-world deployment. The field needs frameworks that tightly integrate causal inference with RL to improve recovery behavior, generalize to unforeseen conditions, and close the gap between theory and practice. Above all, reliance on synthetic data and controlled simulations highlights the need for rigorous validation in realistic driving settings that capture real-world complexity and unpredictability.

\section{Proposed methodology}\label{sec:approach}

\begin{figure*}[tb]
\centering
\includegraphics[width=1.0\textwidth,trim=5pt 20pt 5pt 7pt,clip]{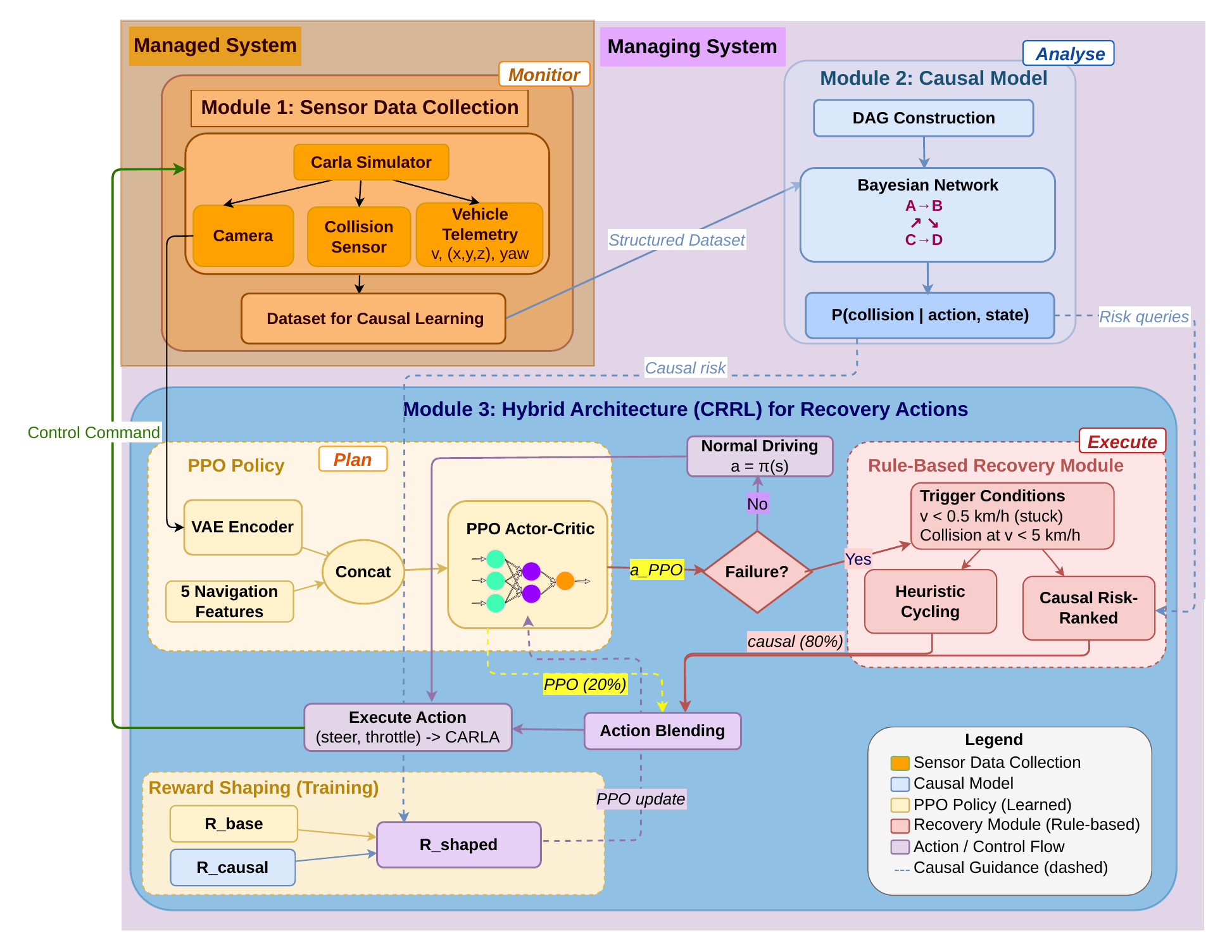}
\vspace*{-2ex}
\caption{Overview of the proposed causal-guided hybrid RL framework, structured as a \textbf{Managing System} operating over the CARLA vehicle environment (\textbf{Managed System}). The three modules follow the MAPE-K feedback loop: (1)~\textit{Monitor}: sensor data collection from the CARLA environment; (2)~\textit{Analyse}: causal model construction via a Bayesian network DAG; and (3)~\textit{Plan/Execute}: a causal-informed PPO policy (Plan) operating alongside a rule-based recovery module (Execute).}
\vspace*{-2ex}
\label{fig:framework}
\end{figure*}

Our proposed methodology combines causal inference and reinforcement learning with a rule-based module within a \textbf{hybrid architecture} to develop a robust and efficient system for detecting and recovering from failures in autonomous driving vehicles. 
The framework employs a three-stage modular design that integrates trajectory data collection, causal model construction, and causal-informed policy training, as illustrated in Figure~\ref{fig:framework}.

The main components of the methodology follow the MAPE-K feedback loop~\cite{iglesia2015:mapek}. \textbf{(1) Sensor Data Collection (Monitor)}: This module continuously monitors the system's state and environmental conditions by logging data from multiple sensors(Figure~\ref{fig:framework} Module~1),  %
\textbf{(2) Causal Model (Analyse)}: Based on the logged data from the previous module, this module builds a directed acyclic graph (DAG) representing causal relationships between system variables(Figure~\ref{fig:framework} Module~2), and %
\textbf{(3) Reinforcement Learning for Recovery Actions (Plan/Execute)}: The final module of the framework focuses on RL-based decision making, which is guided by the causal understanding established in Module~2 (Figure~\ref{fig:framework}). The PPO policy \textit{plans} recovery actions informed by the causal reward signal. The rule-based recovery module \textit{executes} interventions when failure states are detected. This Module~3 (Figure~\ref{fig:framework}) implements a hybrid model architecture \method{} that integrates a learned PPO policy with a rule-based recovery module. Recovery action selection is performed through either heuristic cycling or causal risk-ranked selection, as detailed in Sections~\ref{sec:hybrid_model},~\ref{sec:heuristic_cycling}, and~\ref{sec:causal_risk_ranked}.

The methodology addresses a fundamental limitation of standard reinforcement learning approaches that policies trained without awareness of recovery mechanisms cannot cooperate effectively with them at deployment time. 
The main contribution is a training procedure that uses a causal structure graph extracted from driving logs to shape the reward signal, teaching the PPO policy to anticipate stalled states, avoid failures proactively, and navigate effectively following recovery interventions.

\subsection{Sensor Data Collection (Monitor)}
  This process consists of two steps, both shown in Module~1 of Figure~\ref{fig:framework}: the \emph{Autonomous Driving Environment} (the CARLA Simulator together with its on-board sensor suite) and the \emph{Time-Series Driving Trajectories}, which the framework aggregates into the \emph{Dataset for Causal Learning} block shown in the figure.

\head{Autonomous Driving Environment} 
The system operates within the CARLA simulator~\cite{dosovitskiy2017:carla}, where a vehicle is equipped with the following sensor suite:
\begin{enumerate*}
\item \emph{Camera:} The system employs a front-facing camera with 160$\times$80 pixel resolution and 125$^\circ$ field of view, mounted at an elevated forward position. 
The camera produces class-labelled pixel observations that are converted to the CityScapes palette format~\cite{cordts2016:cityscapes}. These images serve as the primary visual input to the variational autoencoder (VAE)~\cite{kingma:autoencoding}.

\item \emph{Collision sensor:} 
Mounted to the front of the vehicle, this sensor records the impulse intensity of each collision event. The collision history is used for both episode termination and recovery module activation.

\item \emph{Vehicle telemetry:} 
At each simulation step, the system reads vehicle velocity (computed as $v = \sqrt{v_x^2 + v_y^2 + v_z^2} \times 3.6$\,km/h), location $(x,y,z)$, rotation (yaw), throttle, and steering. Lane-center deviation and heading angle relative to the nearest waypoint are derived from these measurements.
\end{enumerate*}

\head{Time-Series Driving Trajectories}
For causal model construction, the framework collects multi-vehicle driving logs from sessions with two autonomous vehicles operating simultaneously. At each timestep, logs record: (1) per-vehicle state (velocity, throttle, steering, brake, reverse status, position), (2) safety events (collision flags, lane-invasion status), and (3) inter-vehicle features (Euclidean distance, time-to-collision). A preprocessing pipeline discretizes continuous variables into categorical bins, merges vehicle streams by episode and timestep, adds spatial and temporal interaction features, and produces structured datasets for causal directed acyclic graph (DAG) construction and RL training.

\subsection{Causal Model (Analyse)}
This module constructs a directed acyclic graph (DAG) from the logged data collected in the previous step.
The DAG represents causal relationships among system variables, capturing multiple aspects, including how different system components affect each other, how failures propagate through the system, and which interventions are most likely to be effective for recovery.
The DAG is constructed in three stages: an initial structure based on domain knowledge and system specifications, refinement using structure learning algorithms applied to observational data, and validation through interventional testing. The generated graph is used in the next step. Details are provided in Section~\ref{sec:causal_graph}.

\subsection{Hybrid RL for Recovery Actions (Plan/Execute)}\label{sec:hybrid_model}

This module extracts state information from the time-series driving trajectories, applies reinforcement learning to determine recovery actions based on the DAG, executes these actions, and assesses failure probability to assign rewards. 
A core contribution of this work is a hybrid architecture that integrates a learned RL policy with a rule-based recovery module, guided by an explicit causal signal. Such hybrid designs mirror practical autonomous systems~\cite{kiran2022:deep}, where learned and engineered components are combined to improve safety and robustness. The proposed decomposition also aligns with the managed/managing system view of self-adaptive systems ~\cite{weyns2012:forms} described in Section~\ref{sec:RW}. The causal guidance and rule-based recovery module serves as the managing subsystem, monitoring and adapting the managed component, specifically the PPO driving policy~\cite{schulman2017:proximal}.%
Prior self-healing work uses RL-based reward-guided adaptation planning for large-scale architectures~\cite {ghahremani2020:improving}. Our work \method{} extends this to physical autonomous vehicles by grounding the reward signal in causal collision-probability estimates. The separation of learned forward driving and rule-based reverse recovery is motivated by two considerations. First, reverse maneuvers are triggered only during rare failure states, making them difficult to learn effectively through RL due to the sparse training signal, and second, rule-based reverse actions provide deterministic, bounded behavior (fixed throttle of $-0.3$), which is preferable for safety-critical recovery, where unpredictable learned actions could worsen the situation. 

The hybrid model consists of three layers:
\begin{enumerate*}[label=\textbf{(\arabic*)}]%
\item \head{Learned Component (PPO Policy)} Responsible for forward driving, obstacle avoidance, recovery-context steering/throttle adjustments, and proactive stalled-state anticipation. The details are discussed in Section~\ref{sec:PPO} 

\item \head{Rule-Based Recovery Module} 
    The rule-based recovery module provides recovery capability through predefined heuristics. Activation occurs when any of the three failure conditions is detected.
    Upon activation, the module selects from a velocity-dependent pool of candidate actions (details in Section~\ref{sec:velocity_candidates}).
    Action selection uses either \textit{heuristic cycling} (Section~\ref{sec:heuristic_cycling}) or \textit{causal risk-ranked selection} (Section~\ref{sec:causal_risk_ranked}). When a forward action is selected during recovery, it is blended with the PPO policy output (80\% recovery guidance, 20\% PPO exploration) reverse actions are applied directly (Algorithm~\ref{algo:training}, lines 5--9)

\item \head{Causal Guidance (Bayesian Network)} 
    The causal guidance layer fulfills two roles within the hybrid architecture. 
    During training, it provides reward shaping that penalizes actions predicted to have high collision probability and rewards low-risk recovery maneuvers. 
    At runtime, when the rule-based recovery module is activated under Condition D, discussed in Table~\ref{tab:ablation_conditions} (causal risk-ranked selection), the causal model computes $P(\text{collision} \mid a, s)$ for each candidate reverse action, enabling the selection of the maneuver with the lowest predicted failure risk.
    This dual use of causal inference, both shaping the learned policy and guiding rule-based decisions, distinguishes the proposed framework from purely learned or purely rule-based approaches.
\end{enumerate*}
The key innovation is that \method{} during training teaches the RL policy to work effectively \textit{with} the recovery module. The four-condition ablation study (Section~\ref{sec:ablation}) demonstrates that this training-time causal guidance is the primary source of improvement. The recovery module alone, without a cooperatively-trained policy, is counterproductive.

\begin{algorithm}[tb]
    \caption{Causal-Guided Hybrid RL Training}
    \label{algo:training}\small
    \begin{algorithmic}
        \State Initialize PPO policy $\pi_\theta$ from pre-trained weights
        \State Load causal Bayesian network $\mathcal{G}$
        \State Initialize recovery module $\mathcal{R}$ with action candidates
        \For{each training episode}
            \State $s_0 \leftarrow \text{env.reset()}$
            \For{each timestep $t$ in episode}
                \If{$\mathcal{R}$.\textsc{ShouldTrigger}$(s_t)$}
                    \State $a_{\text{causal}} \leftarrow \mathcal{R}$.\textsc{SelectAction}$(s_t, \mathcal{G})$
                    \If{$a_{\text{causal}}$ is reverse} \Comment{negative throttle}
                        \State $a_t \leftarrow a_{\text{causal}}$
                    \Else
                        \State $a_{\text{PPO}} \leftarrow \pi_\theta(s_t)$ \Comment{exploration}
                        \State $a_t \leftarrow 0.8 \cdot a_{\text{causal}} + 0.2 \cdot a_{\text{PPO}}$
                    \EndIf
                    \State $r_{\text{causal}} \leftarrow$ \textsc{CausalReward}$(a_t, s_t, \mathcal{G})$
                    \State $r_t \leftarrow r_{\text{task}} + r_{\text{causal}}$
                \Else
                    \State $a_t \leftarrow \pi_\theta(s_t)$ \Comment{normal driving}
                    \State $r_t \leftarrow r_{\text{task}}$
                \EndIf
                \State $s_{t+1} \leftarrow \text{env.step}(a_t)$
                \State Store $(s_t, a_t, r_t, s_{t+1})$ in buffer
            \EndFor
            \State Update $\pi_\theta$ using PPO gradient ascent
        \EndFor
    \end{algorithmic}
\end{algorithm}

The details of each part of the RL module are:
\begin{enumerate*}[label=\textbf{(\arabic*)}]
\item \textbf{State Representation}:
The system represents its state as a 100-dimensional vector combining a 95-dimensional latent encoding from a pretrained VAE applied to the front-facing camera image, concatenated with 5 navigation features (throttle, velocity, normalized velocity, lane-centre distance, and heading angle) shown in PPO Policy input in Figure~\ref{fig:framework}. Full details of the encoding pipeline are provided in Section~\ref{sec:pretrained}.

\item \textbf{Deep RL Policy}: The Reinforcement Learning component is trained to select optimal recovery actions based on the current system state when a failure is detected. The Deep RL policy (i) takes the current state as input, (ii) integrates causal understanding to improve learning efficiency (Section~\ref{sec:causal-integration}), (iii) uses a reward function designed to encourage successful recovery while minimizing intervention costs, and (iv) is trained using simulated environments to generate diverse failure scenarios. 

\item \textbf{Action Selection}: The system combines reinforcement learning with causal inference using the causal model to estimate the effects of candidate actions (Section~\ref{sec:causal_risk_ranked}) to formulate targeted recovery strategies.
During normal driving, the PPO policy selects actions directly as illustrated in the Rule-based recovery module in Figure~\ref{fig:framework}. Upon failure detection (Section~\ref{sec:failure}), the recovery module takes control using the action selection strategies described in Sections~\ref{sec:heuristic_cycling} and~\ref{sec:causal_risk_ranked}.

\item \textbf{Recovery Action}: When a failure is detected, the recovery module assumes control and selects a recovery action using either heuristic cycling or causal risk-ranked selection (Sections~\ref{sec:heuristic_cycling} and~\ref{sec:causal_risk_ranked}). For forward recovery actions, the selected action is blended with the PPO policy output (80\% causal guidance, 20\% PPO exploration), while reverse actions use the recovery module's selection directly. The PPO policy learns from both normal and recovery transitions, enabling it to develop cooperative behaviours over time. The full training procedure is described in Algorithm~\ref{algo:training}.

\item \textbf{Failure Probability Assessment}: The '(failure?)' decision point in Figure~\ref{fig:framework} evaluates whether the recovery action has effectively lowered the failure probability. If yes, the system resumes normal operation, otherwise, it implements additional recovery measures.

\item \textbf{Reward Mechanism}: Recovery action effectiveness is evaluated using the multi-modal reward architecture detailed in Section~\ref{sec:reward} and shown in 'Reward shaping(training)' in Figure~\ref{fig:framework}, which combines base task performance with causal risk shaping to guide the policy towards safer recovery actions.

 \end{enumerate*}

\subsection{Real-time Failure Detection and Recovery} \label{sec:failure} \label{sec:velocity_candidates} 
This step involves testing the proposed framework. During operation, the system continuously monitors for failures and initiates recovery procedures when necessary.

\head{Failure Detection} The system monitors three trigger conditions in real-time:
\begin{enumerate*}[label=\emph{(\arabic*)}]
    \item \textit{Collision:} Any collision event detected by the collision sensor triggers immediate recovery activation.
    \item \textit{Off-road:} Recovery is triggered when the vehicle is not on a driving lane (determined via CARLA waypoint query) or when its distance to the nearest road waypoint exceeds 2.0 m.
    \item \textit{Stalled state:} When velocity remains below 0.5\,km/h for more than 50 consecutive steps, the vehicle is classified as stalled and recovery is triggered. The stalled counter resets once velocity exceeds 2.0\,km/h.
\end{enumerate*}
As a precaution against infinite recovery loops, a limit has been set on the maximum number of recoveries per episode.

\head{Velocity-Dependent Candidate Selection} When recovery is triggered, the system builds a pool of candidate actions based on the current speed of the vehicle:
\begin{enumerate*}[label=\emph{(\alph*)}]
    \item \textit{Very slow} ($v < 1.0$\,km/h): Six candidates including three reverse actions $\{(0.0, -0.3),\; (-0.3, -0.3),\; (0.3, -0.3)\}$ and three forward actions $\{(0.0, 0.4),\; (-0.4, 0.35),\; (0.4, 0.35)\}$.
    \item \textit{Slow} ($1.0 \leq v < 5.0$\,km/h): Four candidates including three forward with varying steering plus a light reverse $(0.0, -0.2)$.
    \item \textit{Moving} ($v \geq 5.0$\,km/h): Three forward-only candidates with gentle steering corrections.
\end{enumerate*}

\head{Recovery Action Selection} 
An action is selected from the candidate pool using one of two strategies, depending on the ablation condition:
\begin{enumerate*}[label=\emph{(\arabic*)}]
    \item \textit{Heuristic cycling} (Conditions B and C, Section~\ref{sec:heuristic_cycling}): The system cycles through candidates in round-robin order, incrementing an index at each recovery step.
    \item \textit{Causal risk-ranked selection} (Condition D, Section~\ref{sec:causal_risk_ranked}): Each candidate is scored by querying the causal Bayesian network for $P(\text{collision} \mid a_i, s_t)$, and the action with the lowest predicted risk is selected. If all causal queries fail, the system defaults to the first candidate in the pool.
\end{enumerate*}

\head{Action Execution} 
The selected action is applied to the vehicle as a (steering, throttle) control command. Negative throttle values engage the reverse gear. The action is applied with the same smoothing as normal driving (exponential blending with the previous steering and throttle values).

\section{Empirical Evaluation}\label{sec:setup}

We define five research questions to systematically evaluate each component of \method{}:
\begin{enumerate}[label=\textbf{RQ\arabic*:},left=2em]
    \item What is the baseline performance of vanilla RL driving without any recovery mechanism?
    \item Does the hybrid system (causal-trained policy + recovery module) outperform the vanilla baseline?
    \item How much external assistance does the recovery module provide?
    \item Does causal-guided training genuinely improve driving behavior, independent of the recovery module?
    \item What are the collision trade-offs of recovery-enabled driving?
\end{enumerate}
RQ1 measures baseline RL policy failure without recovery, establishing intervention necessity. RQ2 evaluates whether the complete hybrid system improves over baseline. Since RQ2 combines causal training with recovery assistance, RQ3 measures recovery intervention frequency to distinguish genuine policy improvement from external dependence. RQ4 tests our central claim by comparing policy quality independent of recovery assistance. RQ5 examines the safety trade-off of increased collision rate and whether performance gains justify this cost. Together, these questions establish baseline performance, isolate component contributions, and assess overall benefits and safety trade-offs. 

\begin{table}[t]
\tablefontsize
\centering
\caption{Four ablation conditions evaluated across three driving scenarios (straight, roundabout, T-junction) with 20 episodes each.}
\label{tab:ablation_conditions}
  \vspace*{-2ex}
\begin{tabular}{clll}
\toprule
\textbf{Condition} & \textbf{Policy Weights} & \textbf{Recovery Module} & \textbf{Purpose} \\
\midrule
A & Pretrained PPO & None & Pure baseline \\
B & Pretrained PPO & Heuristic cycling & Module alone \\
C & Causal-trained PPO & Heuristic cycling & Training effect \\
D & Causal-trained PPO & Causal risk-ranked & Full system \\
\bottomrule
\end{tabular}%
  \vspace*{-2ex}
\end{table}

\subsection{Four-Condition Ablation Evaluation Framework}\label{sec:ablation}

To systematically evaluate the contribution of each component in the proposed hybrid architecture, we conduct a four-condition ablation study that independently varies the policy training approach (pretrained baseline vs. causal-guided) and the recovery mechanism (no recovery, heuristic cycling, or causal risk-ranked selection). The conditions are shown in Table~\ref{tab:ablation_conditions} and
the key pairwise comparisons are:        
\begin{itemize}
    \item \textbf{B\,vs\,A} (Recovery module): Does heuristic recovery alone help a pretrained policy?
    \item \textbf{C\,vs\,B} (Causal training): Does causal-guided training improve the policy, given identical recovery infrastructure?
    \item \textbf{D\,vs\,C} (Causal inference at runtime): Does causal risk-ranked selection outperform heuristic cycling?
    \item \textbf{D\,vs\,A} (Full system): What is the overall improvement of the complete hybrid system?
\end{itemize}
The C\,vs\,B comparison is the central test of our main contribution as both conditions share identical recovery infrastructure (heuristic cycling), any performance difference can \textit{only} arise from differences in learned policy weights. This directly tests whether causal-guided training teaches the policy to cooperate more effectively with recovery interventions.
Each condition is evaluated across the three driving scenarios described in Section~\ref{sec:simulation} (straight, roundabout, and T-junction), representing increasing levels of navigational complexity.
Results are analyzed using Mann-Whitney U tests (non-parametric) with Cohen's $d$ effect sizes at significance levels $p < 0.05$ (*), $p < 0.01$ (**), and $p < 0.001$ (***).

\subsection{Evaluation Metrics}
At each recovery step the system evaluates two conditions:
\begin{enumerate*}[label=\textbf{(\alph*)}]
    \item \textbf{Success:} Velocity exceeds the speed threshold (default 5.0\,km/h). The recovery is logged as successful and normal PPO-driven operation resumes.
    \item \textbf{Timeout:} The maximum recovery duration (default 30 steps in evaluation) elapsed without reaching the speed threshold. The recovery is logged as failed and control returns to the PPO policy. The system can re-trigger recovery if the vehicle remains stalled.
\end{enumerate*}

\subsection{Simulation}\label{sec:simulation}
\begin{figure*}[t]
    \centering
    \includegraphics[width=.5\textwidth]{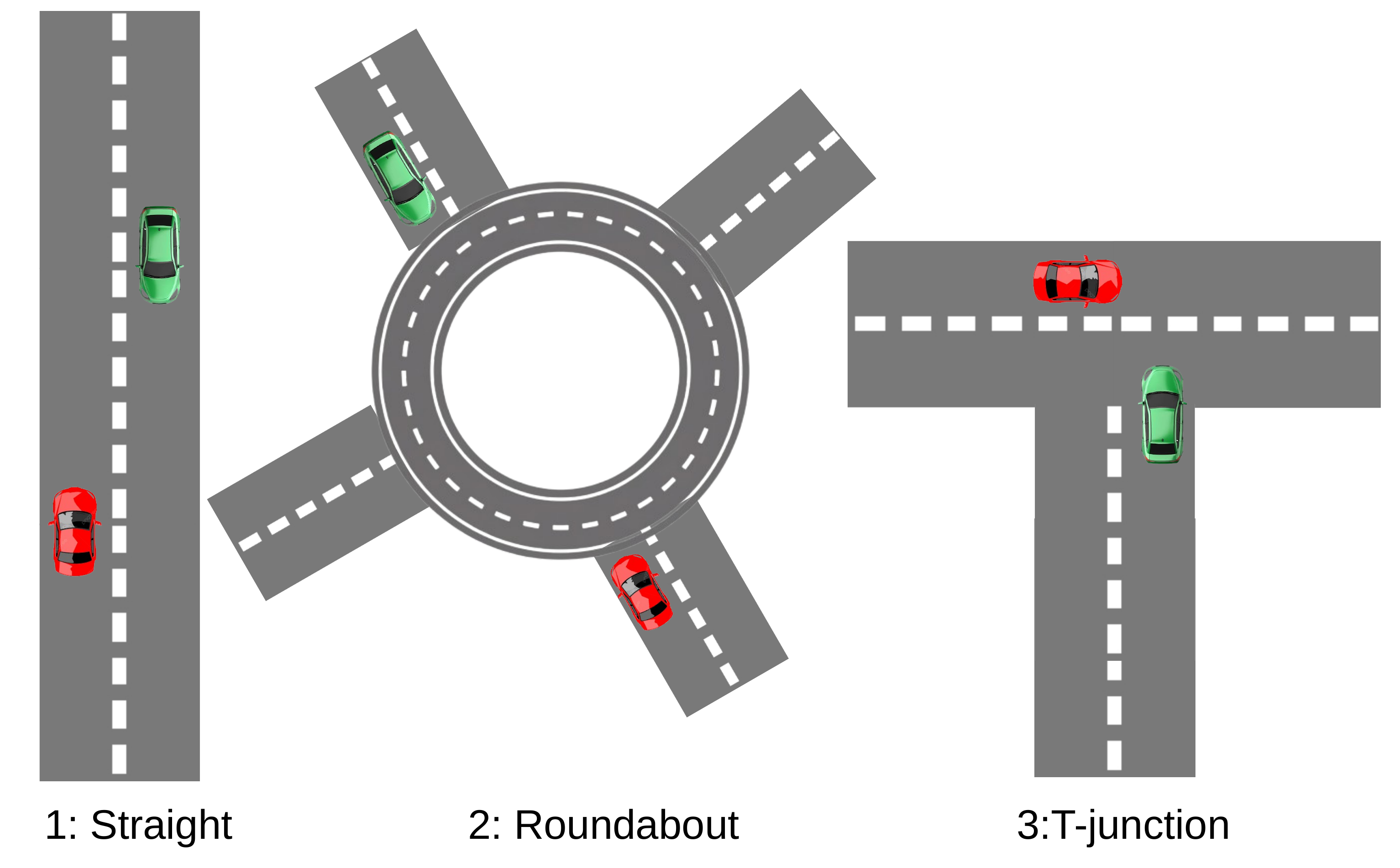}
    \vspace*{-2ex}
    \caption{The three evaluation scenarios.
  (1) Straight road: two vehicles 41\,m apart on opposing lanes.
  (2) Roundabout: vehicles approaching at 165°.
  (3) T-junction: vehicles on perpendicular roads.
  Red car = V1 (pretrained baseline), green car = V2 (CRRL agent).}
		\label{fig:scenario}
  \vspace*{-2ex}
\end{figure*}

We conduct experiments using the CARLA autonomous driving simulator. Training takes place in the Town07 environment, which features
complex urban intersections and multi-lane roads. Two vehicles operate concurrently: Vehicle~1 (V1) employs a pretrained PPO baseline policy, while Vehicle~2 (V2) undergoes online training with causal recovery guidance. To simulate realistic urban traffic, the environment spawns 30 non-player character (NPC) vehicles and 10 pedestrians at randomized locations. Each vehicle is equipped with a camera and a collision sensor for failure detection. We have used three scenarios as shown in Figure~\ref{fig:scenario} to train and evaluate our work.
\begin{enumerate*}[label=\textbf{(\arabic*)}]
    \item \textbf{Straight: }The straight road scenario provides the baseline: two autonomous vehicles in CARLA Town07, facing opposite directions 41 meters apart, maintain lane position at 22 km/h along a 750-meter straight road with pedestrians. This isolates core recovery behaviour without turns or intersections,

    \item \textbf{Roundabout: } The roundabout scenario increases complexity with curved sections and angled intersections. Vehicles approach at 165°, requiring sophisticated steering control. This tests the framework's generalisability to sustained curvature, where lane departure risk is higher, and

    \item \textbf{T-junction:} The T-junction scenario is most challenging. Vehicles on perpendicular approach roads simulate realistic merge situations. This tests recovery under conditions where stalled states and off-road deviations are most likely due to sharp directional changes and complex decision-making.
    
\end{enumerate*} 

\subsubsection{Action and State Space Configuration}
The action and state spaces are detailed in Section~\ref{sec:pretrained}. In summary, the agent uses a 2-dimensional continuous action space (steering and throttle $\in [-1, 1]$). Forward driving maps raw throttle to $[0, \text{max\_throttle}]$ (where $\text{max\_throttle} = 0.4$), while negative values activate reverse gear for recovery maneuvers. Both steering and throttle are exponentially blended with their prior values using a 0.9/0.1 weighting factor to ensure smooth control transitions.

\subsubsection{Data Collection and Preprocessing}
Training data is collected from multi-agent driving episodes in CARLA, each stored as a JSON log. Logs capture agent states and actions per timestep, environmental events (collisions, lane violations), control inputs (steering, throttle, brake, reverse), and episode metadata (map, weather, step limit). These raw states differ from the 100-dimensional latent state representation used by the PPO policy (Section~\ref{sec:pretrained}). The logged data serves as input for causal model construction (Section~\ref{sec:causal_graph}), not for the RL training pipeline directly.

Data preprocessing involves four main transformations: 
(i) feature engineering converts discrete actions into binary indicators (is\_forward, is\_turn, is\_brake), 
(ii) failure events such as collisions and lane violations are combined into binary failure indicators, 
(iii) continuous state variables are discretized using domain-specific thresholds, and 
(iv) timesteps are synchronized across agents through outer joins based on step and episode IDs.

\subsection{Causal Model Architecture} \label{sec:causal_graph}
\begin{figure*}[tb]
\centering
\includegraphics[width=\textwidth]{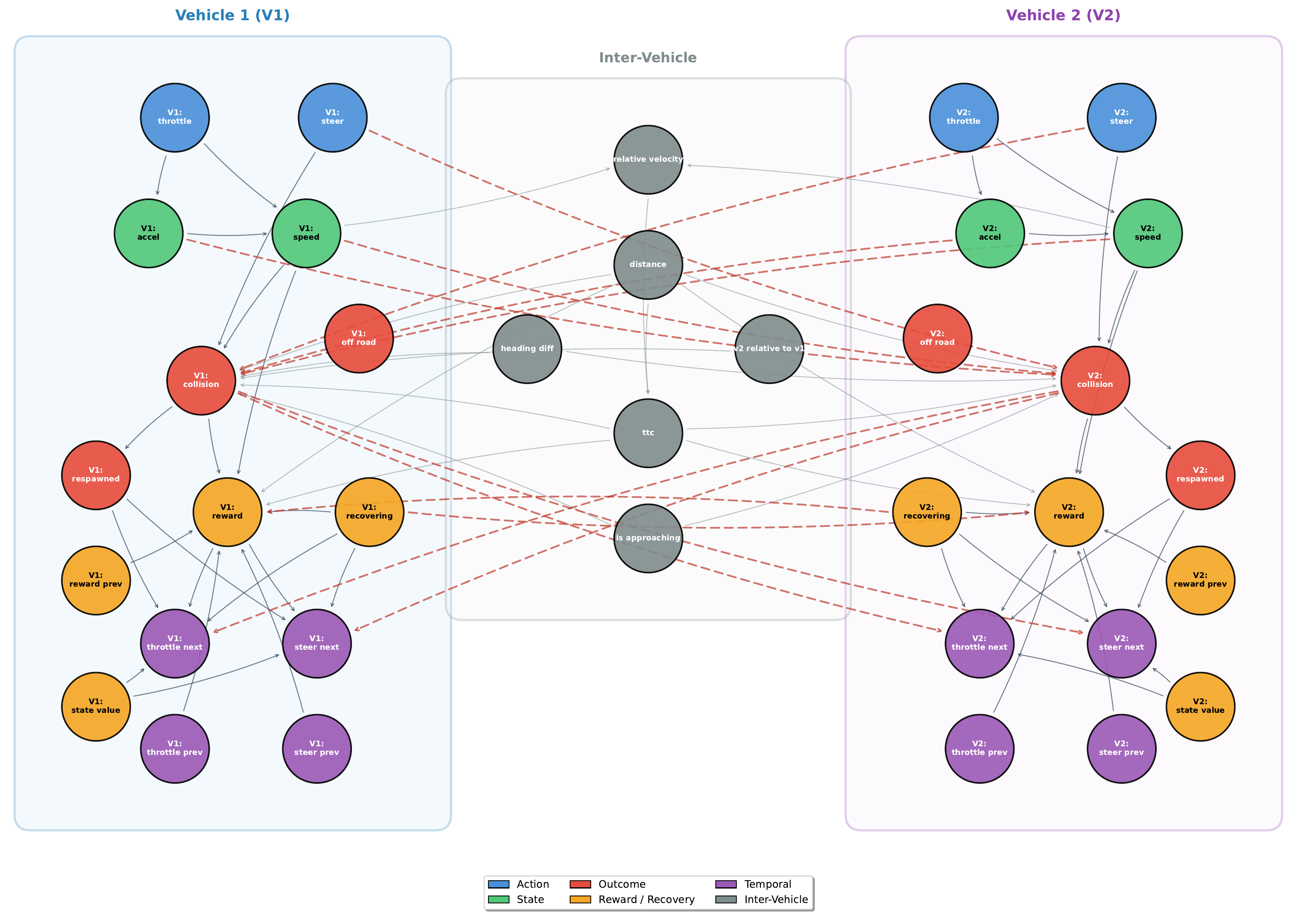}
\vspace*{-2ex}
\caption{Multi-Agent Causal Graph for Two-Car Interaction Scenarios. Node colours indicate category: action (blue), state (green), outcome (red), reward/recovery (orange), temporal (purple), and inter-vehicle spatial (grey). Solid arrows denote intra-vehicle edges; dashed red arrows denote cross-vehicle edges; grey arrows denote spatial-node edges.}
\label{fig:multi_agent_causal_graph}
\vspace*{-1ex}
\end{figure*}

We have implemented the causal model as a Bayesian network built on a DAG using the CausalNex library. Continuous vehicle signals are discretized into categorical bins to serve as graph nodes. Figure~\ref{fig:multi_agent_causal_graph} illustrates the complete causal graph for two-vehicle interaction scenarios.

\subsubsection{Node Categories}

Each vehicle $i \in \{1,2\}$ maintains five categories of nodes:

\begin{enumerate}
\item \textbf{Action nodes} (blue): throttle\_cat (no\_throttle / medium / high) and steer\_cat (hard\_left / left / straight / right / hard\_right). These represent the discretised continuous control commands.

\item \textbf{State nodes} (green): speed\_cat (stopped / slow / medium / fast) and accel\_cat (hard\_brake / decel / constant / accel / hard\_accel). 
These capture the dynamic state of the vehicle.

\item \textbf{Outcome nodes} (red): collision (binary), off\_road (binary), and respawned (binary). These indicate safety violations and post-failure reset events. Collision events from vehicles, pedestrians, other objects are merged into a single binary indicator.

\item \textbf{Reward and recovery nodes} (orange): reward\_cat (negative / zero / positive / high), reward\_prev, state\_value (low / medium / high), and recovering (binary). 
These model the RL feedback loop.

\item \textbf{Temporal nodes} (purple): throttle\_next, steer\_next, throttle\_prev, and steer\_prev. These describe how previous actions affect present states and how present decisions influence subsequent actions.
\end{enumerate}
In addition, six \textbf{inter-vehicle spatial nodes} (grey) are shared between both agents: distance\_cat (critical / close / near / moderate / far), ttc\_cat (imminent / urgent / warning / safe / no\_risk), relative\_velocity\_cat, heading\_diff\_cat, v2\_relative\_to\_v1 (ahead / behind / beside), and \texttt{is\_approaching} (binary).

\subsubsection{Edge Structure}
The causal graph edges are constructed through an expert-guided approach that incorporates domain-specific knowledge of autonomous driving systems. These edges are organized into four categories:

\begin{enumerate}
\item \textbf{Action-to-state and state-to-outcome edges.}
For each vehicle, throttle influences both acceleration and speed (throttle $\to$ accel $\to$ speed, and throttle $\to$ speed directly). 
Speed and steering then determine collision risk (speed $\to$ collision, steer $\to$ collision). Speed also feeds into the reward signal (speed $\to$ reward).

\item \textbf{Reward feedback and value edges.}
The graph captures how the RL learning loop operates over time. 
Current rewards influence subsequent actions (reward $\to$ throttle\_next, reward $\to$ steer\_next), while previous actions affect the present reward (throttle\_prev $\to$ reward, steer\_prev $\to$ reward). 
Rewards propagate across timesteps (reward\_prev $\to$ reward), and the critic's value function guides future control decisions (state\_value $\to$ throttle\_next, state\_value $\to$ steer\_next).

\item \textbf{Recovery and failure-propagation edges.}
Collision events trigger two downstream effects: they feed into the reward signal (collision $\to$ reward) and activate a respawn (collision $\to$ respawned). 
Both the respawn flag and the recovery flag influence the next action (respawned $\to$ throttle\_next, respawned $\to$ steer\_next; recovering $\to$ throttle\_next, recovering $\to$ steer\_next). 
The recovery state also affects the reward (recovering $\to$ reward).

\item \textbf{Inter-vehicle and spatial edges.}
Cross-vehicle interactions are modelled in three ways. 
First, each vehicle's speed, steering, and acceleration are linked directly to the collision node of the other vehicle
Second, the shared spatial nodes (distance, TTC, heading difference, relative position, approaching flag) have edges to both vehicles' collision and reward nodes; vehicle speeds feed into relative velocity, which in turn determines TTC (speed $\to$ relative\_velocity $\to$ TTC). 
Third, collision of one vehicle affects the next actions of the other vehicle (consider, V1 collision $\to$ V2 throttle\_next, V2 steer\_next), and recovery state of one vehicle affects the reward of the other (for instance V1 recovering $\to$ V2 reward).
\end{enumerate}

\subsubsection{Parameter Learning and Inference}
Conditional probability distributions are estimated by training the Bayesian network on the merged, discretized driving logs using CausalNex's fit\_node\_states and fit\_cpds functions. 
During inference, the CausalNex InferenceEngine employs the junction-tree algorithm to calculate collision probabilities via marginal inference: $P(\text{collision}=1 \mid \text{action},\, \text{state})$.

To illustrate, consider a vehicle that is currently traveling at a slow speed (speed\_cat = slow) while approaching another vehicle at close range (distance\_cat = close). The recovery module needs to evaluate the candidate action ``left-turn reverse'' (steer\_cat = hard\_left, throttle\_cat = medium). During parameter learning, the network estimates conditional probability tables from the driving logs. For instance, it records across all logged episodes how often collisions occur when a vehicle steers hard left at a slow speed with another vehicle nearby. At inference time, the system queries:
\begin{equation}
P(\text{collision} = 1 \mid \text{steer\_cat} = \text{hard\_left},          
  \text{throttle\_cat} = \text{medium},     
  \text{speed\_cat} = \text{slow},                                                                                
  \text{distance\_cat} = \text{close})
\end{equation}
If this probability is 0.72, and an alternative candidate ``straight reverse'' yields 
\begin{equation}
P(\text{collision} = 1 \mid \text{straight}, \text{medium}, \text{slow}, \text{close}) = 0.35
\end{equation}
the causal risk-ranked selection (Section~\ref{sec:causal_risk_ranked}) chooses straight reverse as the safer recovery action. The conditional probability tables that enable these queries are estimated automatically from the discretized driving logs using CausalNex's parameter fitting functions.

\subsection{Reinforcement Learning for Recovery Actions}\label{sec:causal-integration}
Our approach integrates causal reasoning with reinforcement learning through a dual-component architecture (Figure~\ref{fig:framework}, Module~3). The causal model serves as both an action evaluator and a reward shaper. Traditional RL methods rely solely on environmental feedback to guide learning. Our system uses causal inference to provide immediate action-level guidance and enhanced learning signals, enabling the policy to anticipate and avoid failure states proactively.

\subsubsection{Pretrained Policy Network}\label{sec:pretrained}
The base driving policy is a Proximal Policy Optimization (PPO)~\cite{schulman2017:proximal} agent pretrained on the standard CARLA Town07 driving task without recovery. The observation pipeline has two parts. First, a Variational Autoencoder (VAE) compresses front-facing RGB camera images ($160\times80\times3$) into a 95-dimensional latent representation. The encoder consists of four convolutional layers (3$\to$32$\to$64$\to$128$\to$256 channels) with LeakyReLU activations and batch normalisation, followed by fully connected layers mapping the flattened feature map (9,216 dimensions) to a 1,024-dimensional intermediate representation, which is then projected to a 95-dimensional latent space via the reparameterisation trick. The VAE is trained separately on driving data and remains frozen throughout subsequent RL training, solely serving as a fixed perceptual feature extractor. Second, a 5-dimensional navigation vector captures the current throttle, raw velocity (km/h), normalised velocity ($v / v_{\text{target}}$), normalised distance from lane centre, and normalised heading angle. These are concatenated to form a 100-dimensional observation vector.

The PPO agent follows an actor-critic architecture. Both actor and critic networks consist of three fully connected hidden layers of sizes 500, 300, and 100 with Tanh activations. The actor outputs a 2-dimensional continuous action vector (steering, throttle) through a final Tanh layer, parameterised as a multivariate Gaussian with diagonal covariance. The critic outputs a scalar state-value estimate. The policy is optimised using the Adam optimiser with a learning rate of $1\times10^{-4}$ for both networks. The action standard deviation is initialised at 0.2 and decayed by 0.05 every 500,000 timesteps to a minimum of 0.05, following a scheduled exploration reduction strategy. Training runs for 2,000,000 timesteps with episodes of up to 7,500 steps each. During original pretraining, the reward function encourages lane-centred driving at a target speed of 22~km/h, computed as the product of a centering factor, an angle alignment factor, and a velocity-dependent scaling term, with penalties of $-10$ applied for collisions, off-road deviations, prolonged stalled states, and overspeeding beyond 25~km/h. Further details of the pretrained model are described in~\cite{razak2023:autonomous}.

In the experimental evaluation, this pretrained policy serves as the shared initialisation for all conditions. Vehicle~1 (baseline) uses the frozen pretrained weights, while Vehicle~2 is fine-tuned with the proposed causal-guided recovery framework, enabling a controlled comparison where both vehicles begin from an identical policy.

\subsubsection{PPO Agent with Causal Integration}\label{sec:PPO}

Building on the pretrained policy described in Section~\ref{sec:pretrained}, the causal-guided agent extends PPO by integrating causal inference at two points within the training loop: \textit{action blending during recovery} and \textit{causal reward shaping}. The network architecture and state representation remain identical to the pretrained model. During causal-guided training, the action standard deviation is fixed at 0.6 to maintain broad exploration, rather than the decaying schedule used during pretraining.

\head{Recovery Trigger} The causal recovery module activates when the agent enters a failure state, defined by one of three boolean conditions: (i) a collision is detected via the collision sensor, (ii) off-road departure is detected using the CARLA waypoint API, where the absence of a valid driving-lane waypoint within 2~m of the vehicle position is treated as off-road, or (iii) a stalled state occurs, defined as velocity below 0.5~km/h persisting for more than 50 consecutive steps. Under normal driving conditions, the agent operates as a standard PPO policy without any causal intervention.

\head{Action Blending} During a recovery phase, both the causal model and the PPO actor generate actions simultaneously at each timestep. The causal model queries the Bayesian network over a velocity-dependent candidate pool (Section~\ref{sec:heuristic_cycling}) and selects the action with the lowest predicted failure probability:   
\begin{equation}                                                                                               a^{*}_{\text{causal}} = \arg\min_{a_i \in \mathcal{A}} \; P(\text{failure}=1 \mid a_i,\, s_t)
\end{equation}                                                                                                                  
The PPO actor independently samples $a_{\text{PPO}} \sim \pi_\theta(\cdot \mid s_t)$ with training enabled, so that the gradient information is preserved. The executed action is then a weighted blend of the two:
\begin{equation}                                                                                               a_t = \alpha \, a^{*}_{\text{causal}} + (1-\alpha)\, a_{\text{PPO}}, \quad \alpha = 0.8
\end{equation}
For reverse maneuvers (throttle $< 0$), the causal action is applied without blending since PPO has no experience with reverse control during pretraining. This design allows the policy to observe and learn from causal guidance while retaining sufficient exploratory variance for gradient-based updates.
  
\head{Reward Shaping} During recovery, the PPO reward stored in memory is increased by a causal bonus proportional to the predicted safety of the executed action.
\begin{equation} 
      r_t^{\text{shaped}} = r_t^{\text{env}} + \bigl(1 - P(\text{failure}=1 \mid a_t,\, s_t)\bigr) \times 0.5                   
\end{equation}                                                                                                                  
This provides dense, per-step feedback that rewards actions which reduce collision risk, even in timesteps where no collision actually occurs. Under sparse environmental rewards alone, the policy receives a signal only upon failure, the causal bonus accelerates credit assignment by continuously differentiating safe from risky actions throughout the recovery trajectory.  Outside recovery phases, $P(\text{failure})$ is not queried, and no shaping is applied.    
  
\head{Policy Update} 
At the end of each episode, the PPO policy is updated using the standard clipped surrogate objective over all experiences collected during that episode, with the causally shaped rewards as the learning signal:
\begin{equation}     
  \begin{split}    
  L^{\text{CLIP}} = &-\min\!\bigl(r_t(\theta)\hat{A}_t,\; \text{clip}(r_t(\theta), 1{-}\epsilon, 1{+}\epsilon)\hat{A}_t\bigr) + 0.5\,L^{\text{VF}} - 0.01\,H[\pi_\theta]                                 \end{split}
\end{equation}
where $r_t(\theta)$ is the probability ratio, $\hat{A}_t$ the advantage estimate computed from the shaped returns, $L^{\text{VF}}$ the value function MSE loss, and $H[\pi_\theta]$ the policy entropy. The full set of hyperparameters are listed in Table~\ref{tab:training_causal integration}.
  
\begin{table}[b]
  \vspace*{-2ex}
\tablefontsize
  \caption{Hyperparameters for PPO with Causal Guidance}                                          \label{tab:training_causal integration}
  \vspace*{-2ex}
  \centering
  \begin{tabular}{ll}
  \begin{tabular}{ll}
  \toprule                     
  \textbf{Parameter} & \textbf{Value} \\
  \midrule                  
  \multicolumn{2}{l}{\textit{Network Architecture}} \\                   
  \midrule
  State dimension & 100 (95 VAE + 5 navigation) \\
  Action dimension & 2 (steering, throttle) \\           
  Hidden layers & 500 $\to$ 300 $\to$ 100 \\
  Activation & Tanh \\                 
  \midrule
  \multicolumn{2}{l}{\textit{Causal Integration}} \\
  \midrule
  Action blend ratio $\alpha$ & 0.8 \\
  Causal reward coefficient & 0.5 \\
  Recovery duration & 30 steps \\
  Recovery success threshold & 5.0~km/h \\
  Stalled trigger threshold & 50 steps \\  
  \bottomrule
  \end{tabular}  
&
  \begin{tabular}{ll}
  \toprule                     
  \textbf{Parameter} & \textbf{Value} \\
  \midrule
  \multicolumn{2}{l}{\textit{PPO Training}} \\                                                             
  \midrule
  Learning rate & $1\times10^{-4}$ \\                                                                        
  Discount factor $\gamma$ & 0.99 \\
  Clip parameter $\epsilon$ & 0.2 \\
  Entropy coefficient & 0.01 \\
  Value function loss coefficient & 0.5 \\
  Updates per iteration & 7 \\
  Initial action std & 0.6 \\
  \bottomrule \\[8.5ex]
  \end{tabular}  
  \end{tabular}  
\end{table}       

\subsubsection{Causal Action Evaluation}
The causal model provides real-time action evaluation by computing collision probabilities for continuous recovery action candidates, as shown in Algorithm~\ref{algo:causal}. 
Each candidate action is represented as a continuous tuple a\_i = (steering\_i, throttle\_i) sampled from velocity-dependent candidate pools defined in Section~\ref{sec:heuristic_cycling}.

For each candidate, the system constructs an evidence vector combining the current vehicle state (velocity, throttle, steering, distance) to other vehicles, and TTC, with action features derived from the continuous tuple. The Bayesian network computes $P(\text{has\_failure} = 1 \mid E)$ via probabilistic inference.
If a causal query fails (for example, due to unseen evidence states), that candidate is skipped. If all queries fail, the system defaults to the first action in the candidate pool.

\begin{algorithm}
\caption{Bayesian Network-based Action Failure Prediction}
\small
\begin{algorithmic}
\Require Set of continuous candidate actions $\mathcal{A} = \{a_1, a_2, \ldots, a_n\}$ where $a_i = (\text{steer}_i, \text{throttle}_i)$
\Require Current vehicle state $S_t = \{v_t, d_t, \theta_t, \ldots\}$ (velocity, distance, heading, etc.)
\Require Trained Bayesian Network $\mathcal{BN}$
\Ensure Failure probabilities $\mathbf{P} = \{p_1, p_2, \ldots, p_n\}$

\State Initialize failure probability vector $\mathbf{P} \leftarrow \emptyset$

\For{each candidate action $a_i = (\text{steer}_i, \text{throttle}_i) \in \mathcal{A}$}
    \State $E \leftarrow \{v_t, d_t, \theta_t, \text{steer}_i, \text{throttle}_i\}$ \Comment{Step 1: Build evidence vector from current state}

    \State $p_i \leftarrow P(\text{has\_failure} = 1 \mid E, \mathcal{BN})$
    \Comment{Step 2: Query Bayesian Network}

    \If{query failed}
    \Comment{Step 3: Handle query failure}
        \State Skip this candidate
    \Else
        \State $\mathbf{P} \leftarrow \mathbf{P} \cup \{p_i\}$ where $p_i \in [0,1]$
    \EndIf
\EndFor

\If{$\mathbf{P} = \emptyset$}
    \State \Return fallback to first candidate with $p = 0.5$
\EndIf

\State \Return $\mathbf{P}$
\end{algorithmic}
\label{algo:causal}
\end{algorithm}

\subsubsection{Multi-Modal Reward Architecture}\label{sec:reward}

The reward function is shaped using causal failure probability estimates from the Bayesian network to guide the PPO policy toward cooperative recovery behaviour~\cite{ng:policy}. The reward system combines environmental outcomes with causal risk predictions. The reward architecture consists of the following components:

\begin{enumerate}
\item \textbf{Velocity reward:} A tiered reward based on the current speed of the vehicle, encouraging forward progress:
\begin{align}
\setlength{\tabcolsep}{15pt}
\begin{tabular}{lccccc}
$v$ (km/h)      & $>15$ & $>10$ & $>5$ & $>2$ & $<0.5$ (stalled) \\
\midrule
$R_{velocity}$  & 1.5   & 1.0   & 0.5  & 0.2  & $-0.3$  \\
\end{tabular}
\end{align}

\item \textbf{Collision penalty:} A flat penalty applied when a collision is detected:
\begin{align}
R_{collision} = \begin{cases}
-20.0 & \text{if collision occurred} \\
0 & \text{otherwise}
\end{cases}
\end{align}

\item \textbf{Recovery progress bonus:} An additional reward for maintaining forward motion during active recovery, encouraging the vehicle to break free from stalled states:
\begin{align}
R_{recovery} = \begin{cases}
1.0 & \text{if in recovery and } v > 3.0 \text{\,km/h} \\
0 & \text{otherwise}
\end{cases}
\end{align}

\item \textbf{Causal risk shaping:} During recovery, when the causal model is available and the vehicle has regained some motion ($v > 1.0$\,km/h), a bonus inversely proportional to predicted collision risk is applied:
\begin{align}
R_{causal}(s_t, a_t) = (1.0 - P(\text{failure} \mid s_t, a_t)) \times 0.5
\end{align}
This guides the agent towards lower-risk recovery actions as assessed by the causal Bayesian network.

\end{enumerate}
During recovery episodes, the shaped reward adds the causal component to the base environmental reward:
\begin{align}
R_{shaped} = R_{base} + R_{causal}
\end{align}
where $R_{base} = R_{velocity} + R_{collision} + R_{recovery}$. During normal (non-recovery) driving, the reward is simply $R_{base}$.

This reward architecture encourages forward progress at all times, strongly penalises collisions, provides additional incentive during recovery to regain motion, and uses the causal model's risk predictions to steer the policy towards safer recovery actions.

\subsubsection{Recovery Mode:}
Our system operates in two distinct modes. In \textbf{Normal Operation} mode, the PPO policy $\pi_\theta$ selects actions directly from the state representation. The reward is purely environmental ($R_{base}$) without causal shaping, while the system continuously monitors for failure conditions. \textbf{Recovery Mode} is activated when any of the three failure conditions described in Section~\ref{sec:failure} are detected. Once activated, the recovery module assumes control and selects actions via the hybrid model architecture (Section~\ref{sec:hybrid_model}). The reward switches to the shaped formulation $R_{shaped}$ (Section~\ref{sec:reward}). Recovery termination conditions are defined in Section~\ref{sec:velocity_candidates}.

\subsubsection{Heuristic Cycling}\label{sec:heuristic_cycling}
Heuristic cycling employs a deterministic round-robin approach to choose recovery actions from candidate pools that vary with velocity. 
This mechanism provides a straightforward baseline for recovery and is applied in ablation Conditions B and C (Section~\ref{sec:ablation}).

When recovery is triggered, the system first selects a candidate pool based on the current velocity of the vehicle (see Section~\ref{sec:velocity_candidates}), then cycles through the candidates in round-robin order:
$ %
    a_t = \mathcal{A}_{\text{candidates}}[k \bmod |\mathcal{A}_{\text{candidates}}|], \quad k \leftarrow k + 1
$ %
where $k$ is a persistent counter incremented at each recovery step. This ensures different actions are tried systematically, regardless of proximity to obstacles or collision risk.

The key properties of heuristic cycling are
\begin{enumerate*}[label=\textbf{(\alph*)}]
    \item \textbf{Velocity-aware but not risk-aware} The candidate pool adapts to speed, but no probabilistic risk assessment guides the selection within the pool,
    \item \textbf{Deterministic} Given the same velocity range, the mechanism produces an identical action sequence on every invocation,%
    \item \textbf{No collision prediction} The system cannot anticipate whether a candidate action will result in a collision, and
    \item \textbf{Ablation baseline} Provides equivalent recovery \textit{capability} across conditions. this approach isolates the effect of causal training on policy weights in the C vs B comparison.
\end{enumerate*}

\subsubsection{Causal Risk-Ranked Selection}\label{sec:causal_risk_ranked}
Causal risk-ranked selection is an intelligent, state-aware strategy to choose recovery actions. It uses the causal Bayesian network to predict collision risk for each candidate action and selects the safest option. This mechanism is employed in ablation Condition D and represents the full system capability.
\head{Risk Prediction}
For each candidate recovery action $a_i \in \mathcal{A}_{\text{candidates}}$, the causal model predicts the probability of collision:
  $  r_i = P(\text{collision} = 1 \mid a_i, s_t)$
where $s_t$ includes velocity (km/h), distance to other vehicles (m), and time-to-collision (s). The action with the lowest predicted risk is selected:
  $  a^* = \arg\min_{a_i \in \mathcal{A}_{\text{candidates}}} r_i$

\head{Fallback Mechanism}
When all causal queries fail (Foe example, due to missing evidence states in the Bayesian network), the system defaults to the first candidate, ensuring recovery always produces an action. 
To improve coverage, the system cycles through subsets of candidates across recovery steps, evaluating up to three candidates per step.

\head{Recovery Termination}
Recovery termination follows the exit conditions defined in Section~\ref{sec:velocity_candidates} that is success when velocity exceeds 5.0~km/h, timeout after 30 steps, and a per-episode cap of 5 successful recoveries.

\head{Comparison of Selection Strategies}
Table~\ref{tab:selection_comparison} compares the two recovery action selection strategies.

\begin{table}[b]
\tablefontsize
\centering
\caption{Comparison of the two recovery action selection strategies.}
\label{tab:selection_comparison}
\vspace*{-2ex}
\begin{tabular}{lccccc} 
\toprule
\thead{Strategy} & \thead{State-aware} & \thead{Risk\\assessment} & \thead{Collision\\prediction} & \thead{Fallback on\\query failure} & \thead{Computational\\cost}  \\ 
\midrule
Heuristic Cycling            & No                              & No                                  & No                                       & N/A                                           & Minimal                                 \\ 
Causal Risk-Ranked           & Yes                             & Yes                                 & Yes                                      & Yes                                           & Moderate                                \\
\bottomrule
\end{tabular}
\vspace*{-1ex}
\end{table}

\subsubsection{Causal-Guided Action Selection Process}

The action selection process follows a two-phase approach, summarised in Algorithm~\ref{algo:training}. During \textbf{training}, the PPO policy selects actions during normal driving. When recovery is triggered (Section~\ref{sec:failure}), the action blending mechanism described in Section~\ref{sec:hybrid_model} applies: 80\% causal guidance with 20\% PPO exploration for forward actions, and pure causal actions for reverse maneuvers. The PPO agent learns from both normal and recovery transitions, with rewards shaped by the causal component during recovery (Section~\ref{sec:reward}).

During \textbf{evaluation}, the identical state encoding pipeline is employed. The PPO policy selects actions during normal driving. Upon detection of recovery triggers, the recovery module takes control and selects actions through either heuristic cycling (Conditions~B, C) or causal risk-ranked selection (Condition~D). The PPO policy does not train during evaluation; its weights remain fixed from the training phase.

\subsubsection{Causal Model Integration Architecture}
The causal model functions as an independent advisory module that operates in parallel with the PPO policy. The integration occurs through two primary interfaces: (i) reward shaping during recovery, as defined in Section~\ref{sec:reward}, and (ii) direct recovery action selection in Condition~D via causal risk-ranked selection (Section~\ref{sec:causal_risk_ranked}).
If a causal query fails for a given candidate (for example, due to unseen evidence combinations in the Bayesian network), that candidate is skipped and the system defaults to the first candidate in the pool if all queries fail.

\subsubsection{Learning Dynamics and Convergence}
Causal-guided training modifies standard PPO learning through two mechanisms. First, the shaped reward signal (Section~\ref{sec:reward}) provides denser feedback during recovery than sparse environmental rewards alone, accelerating recovery behaviour acquisition. Second, the action blending mechanism (Section~\ref{sec:hybrid_model}) exposes the policy to effective recovery trajectories while preserving exploration via its own action distribution during forward motion.

\subsection{Overall Framework Training}

Training follows a two-stage pipeline. In stage 1 \textbf{causal model fitting}, CARLA driving logs are preprocessed: continuous features are discretized, vehicle data merged, and spatial-temporal features added. A DAG is constructed and a Bayesian network fitted using CausalNex, with timeout protections (60 s structure learning, 120 s fitting).

In stage 2, \textbf{\method{} training}, a two-vehicle experiment runs in CARLA Town07. V1 uses pretrained PPO with no recovery as baseline. V2 starts from the same weights but trains online with causal recovery. The PyTorch implementation uses Adam, performing 7 gradient updates per iteration with Monte Carlo returns. Training runs up to 2000 episodes of 500 steps, with V2's weights saved periodically. Computation runs on CPU, with the frozen VAE optionally on GPU.

\subsubsection{Training Convergence Analysis}\label{sec:training_convergance}

\begin{figure}[tb]
\centering
\begin{minipage}{.49\textwidth}
\centering
\includegraphics[width=\textwidth]{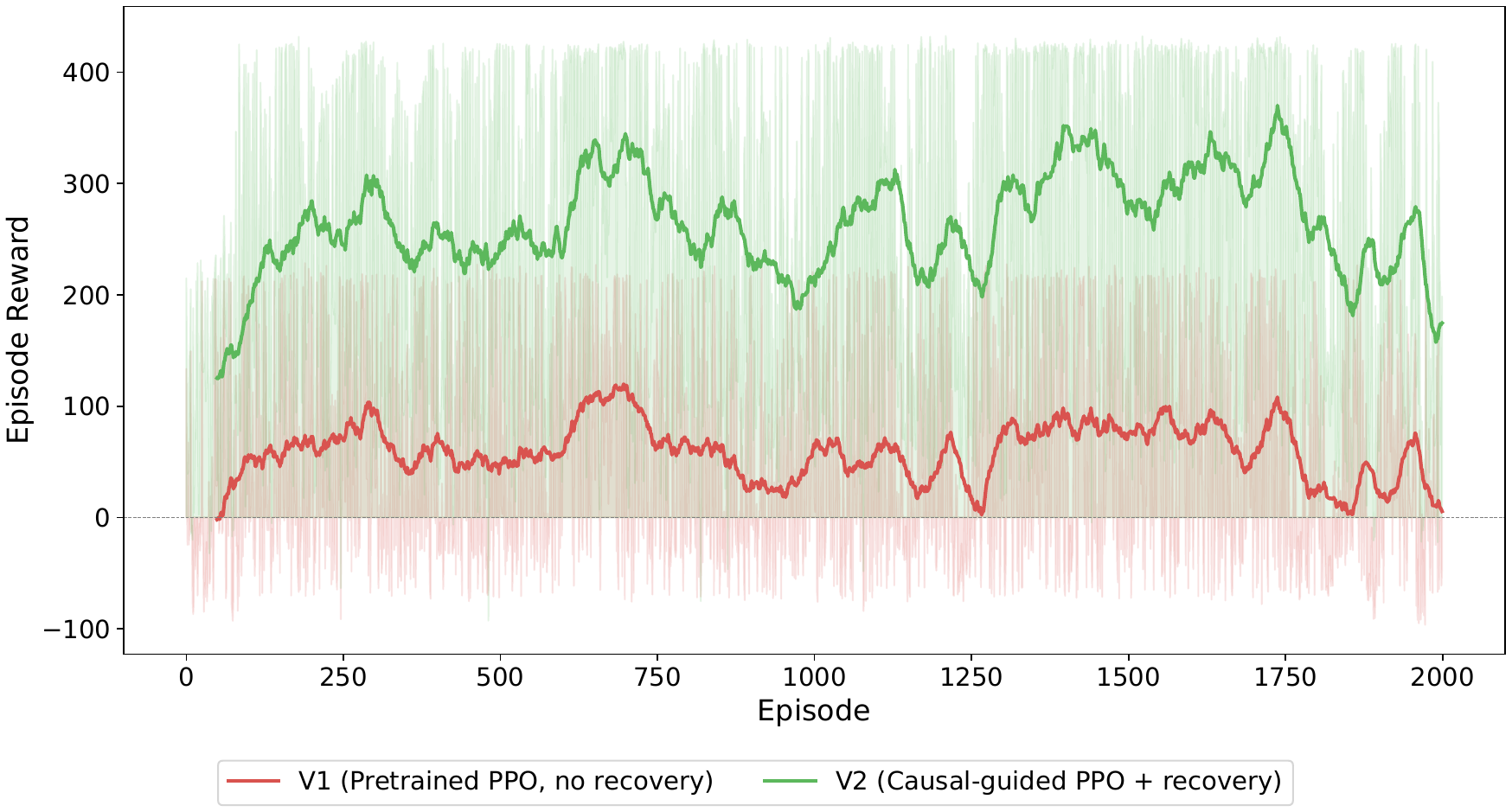}
\vspace*{-2ex}
\caption{Training progress over 2000 episodes (50-episode moving average). V2 (causal-guided PPO with recovery, green) consistently outperforms V1 (pretrained PPO baseline, red) throughout training. Raw per-episode values shown with light shading.}
\label{fig:training_progress}
\end{minipage}
\hfill
\begin{minipage}{.49\textwidth}
\centering
\includegraphics[width=\textwidth]{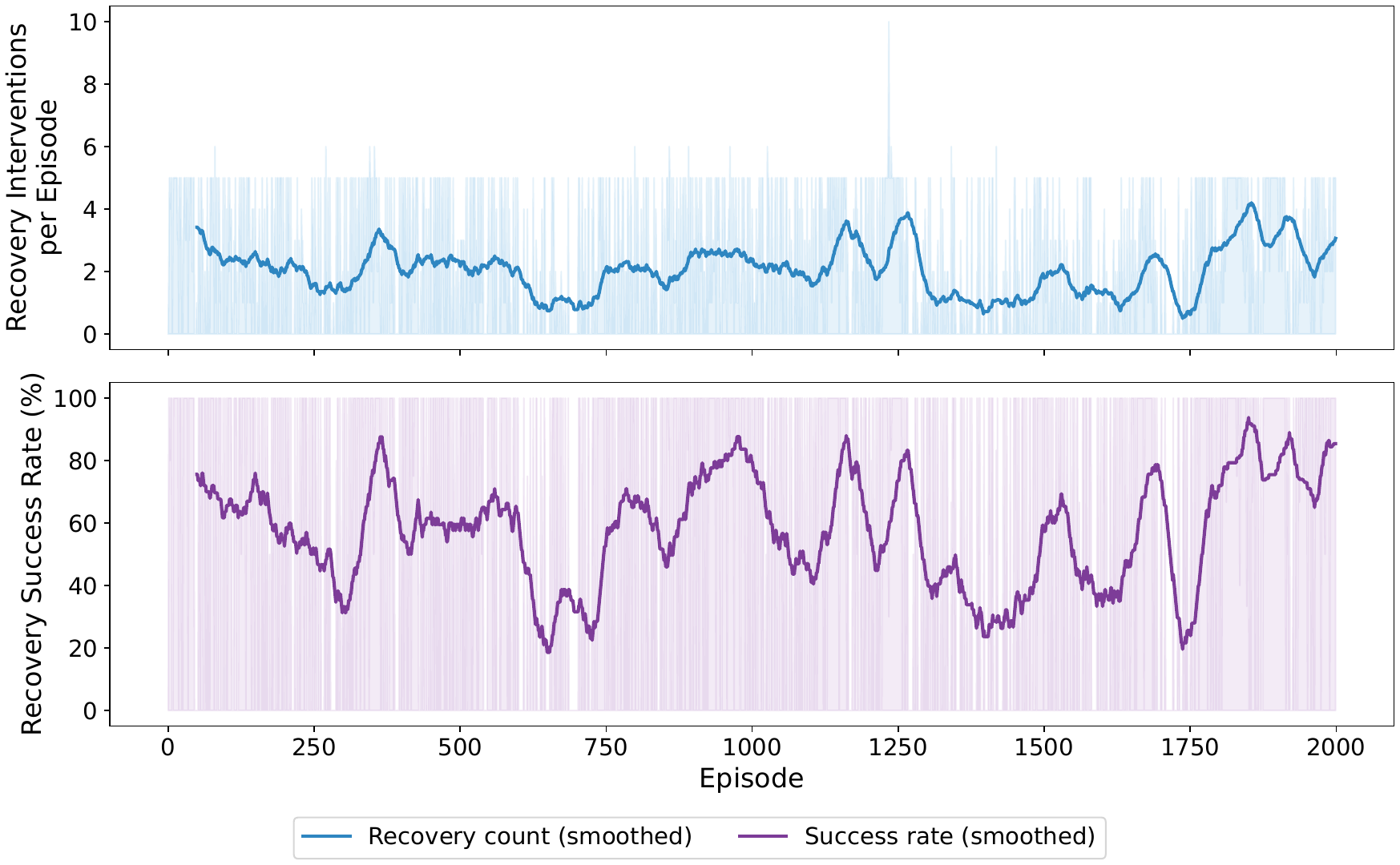}
\vspace*{-2ex}
\caption{V2 Recovery Dynamics During Training (50-episode moving average). \textit{Top:} Number of recovery interventions per episode. \textit{Bottom:} Recovery success rate (\%), averaging 58.2\% with high variance across episodes.}
\label{fig:recovery_dynamics}
\end{minipage}
\vspace*{-2ex}
\end{figure}

The training process runs for 2000 episodes in CARLA Town07, with both Vehicle~1 (V1, pretrained baseline) and Vehicle~2 (V2, causal-guided) operating simultaneously. Figure~\ref{fig:training_progress} shows the episode reward progression for both vehicles over the full training run.

The training dynamics exhibit several key characteristics:
\begin{enumerate*}[label=\textbf{(\arabic*)}]
\item \textbf{Immediate Advantage of Causal Recovery} 
V2 achieves substantially higher rewards from the first episodes. The causal recovery module enables escape from stalled states, yielding a mean episode reward of 260.6 ($\sigma = 139.7$) for V2 versus 57.9 ($\sigma = 97.7$) for V1, which is a $4.5\times$ improvement.
\item \textbf{V2 Policy Refinement} 
The mean reward of V2 improves from $\approx$205 (first 200 episodes) to $\approx$212 (last 200 episodes), showing continued learning of recovery behaviors. The shaped reward signal provides denser feedback during recovery episodes.
\item \textbf{High Variance} 
Both vehicles exhibit substantial variance due to stochastic traffic (30 NPCs, 10 pedestrians) and spawn variability. The 50-episode moving average reveals the underlying trend.%
\end{enumerate*}

\subsubsection{Recovery Dynamics Analysis} \label{sec:Recovery_dynamics}

Figure~\ref{fig:recovery_dynamics} shows the recovery intervention count and success rate for V2  across the training period, revealing how the causal recovery system functions and evolves.
The recovery dynamics reveal key characteristics of the causal-guided training process:

\head{Recovery Frequency} 
During training on the straight-road scenario, V2 averages 2.1 recovery interventions per episode (range 0--10), showing stalled states occur regularly and the recovery module is actively used. Note that evaluation-time recovery counts are higher (5.15--8.70 per episode, Table~\ref{tab:rq3}), reflecting longer evaluation episodes and scenario-dependent complexity. 

\head{Recovery Success Rate} 
The recovery success rate averages 58.2\% across all episodes, with high variability due to the stochastic traffic environment. The 50-episode moving average shows a modest upward trend (from $\approx$67\% to $\approx$82\%), though this requires careful interpretation given the substantial variance. The recovery module maintains a functional success rate throughout training, indicating the PPO policy does not interfere with recovery interventions.

\head{Episodes with Zero Recovery}
A meaningful fraction of V2's episodes require no recovery interventions at all. The zero-recovery analysis in Section~\ref{sec:rq3} (Table~\ref{tab:zero_recovery}) provides further evidence: even in episodes where the recovery module is never triggered, V2 outperforms V1 on distance, reward, and stalled rate, suggesting that causal-guided training teaches the policy to proactively avoid stalled states rather than merely relying on the recovery module to escape them

These results validate the core claim that causal-guided training enables the policy to develop behaviors that work with the rule-based recovery module, achieving higher reward and improving recovery effectiveness.

\section{Results and Discussion}

\subsection{RQ1: Baseline Performance Without Recovery}
To analyze the baseline performance, we have evaluated Condition A (pretrained PPO, no recovery) for all three scenarios using the evaluation metrics shown in Table~\ref{tab:rq1}.
In all three scenarios, the vanilla policy exhibits high stalled rates and limited navigation distance.

\begin{table}[t]
\tablefontsize
\centering
\caption{Condition A baseline performance (vanilla PPO, no recovery).}
\label{tab:rq1}
\vspace*{-2ex}
\begin{tabular}{lccc}
\toprule
\textbf{Metric} & \textbf{Straight} & \textbf{Roundabout} & \textbf{T-Junction} \\
\midrule
Stalled time (\%) & 26.89 & 36.36 & 67.45 \\
Off-road time (\%) & 45.51 & 0.25 & 37.04 \\
Distance (m) & 36.09 & 30.45 & 16.20 \\
Reward & 92.39 & 56.56 & $-$47.95 \\
Collisions & 0.85 & 0.75 & 1.00 \\
\bottomrule
\end{tabular}
\end{table}
Table~\ref{tab:rq1} reveals that without recovery mechanisms, the PPO policy remains stalled for 26–67\% of each episode, with the proportion varying by scenario difficulty. The T-junction presents the greatest challenge: vehicles are immobilized 67.45\% of the time, drive off-road 37.04\% of the time, and achieve a negative cumulative reward of $-$47.95. These findings demonstrate that recovery systems are essential for navigating complex urban environments.

\begin{figure}[tb]
\centering
\includegraphics[width=\textwidth]{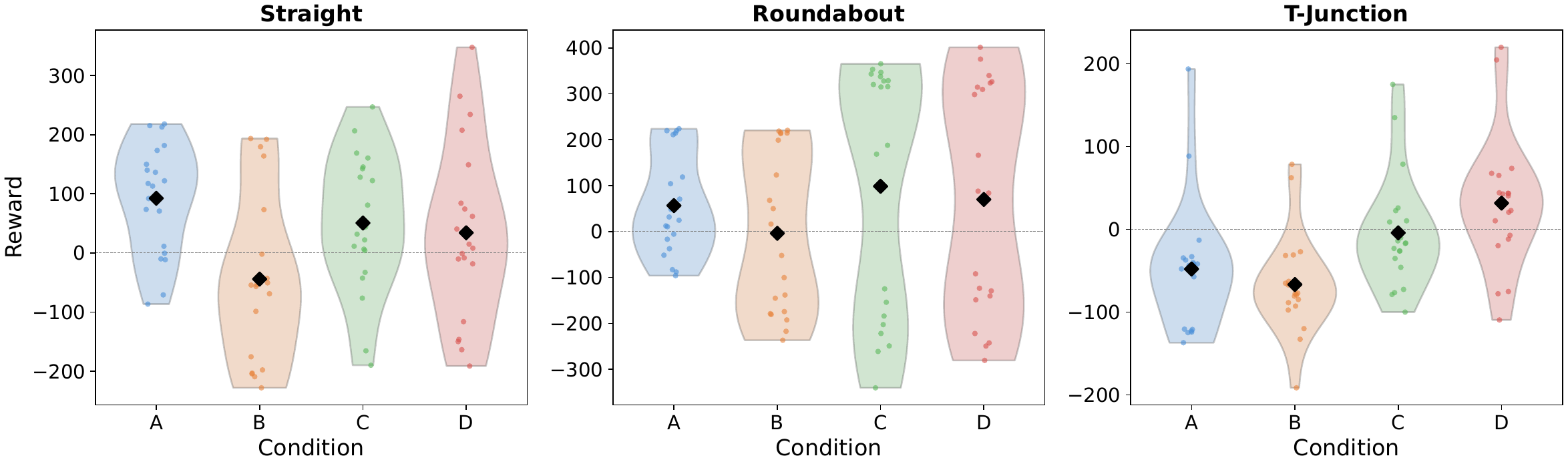}
\vspace*{-2ex}
\caption{Reward distributions across all four ablation conditions described in Table~\ref{tab:ablation_conditions}. Violin plots show individual episode rewards (dots), means (black diamonds), and distributional shape. Conditions C and D consistently shift the distribution upward relative to A and B, particularly in roundabout and T-junction scenarios.}
\label{fig:reward_distributions}
\end{figure}

Figure~\ref{fig:reward_distributions} shows the reward distributions across all four conditions for each scenario. The violin plots reveal that Condition A (blue) exhibits high variance, with many episodes yielding negative rewards, particularly in the roundabout and T-junction scenarios. This variability reflects the unreliability of vanilla PPO in complex urban environments.

\subsection{RQ2: Hybrid System vs. Vanilla Baseline }\label{sec:rq2}%

\begin{table}[b]
\tablefontsize
\centering
\caption{Hybrid system effectiveness: Condition C vs Condition A.}
\label{tab:rq2}
\vspace*{-2ex}
\begin{tabular}{lcccccc}
\toprule
 & \multicolumn{2}{c}{\textbf{Straight}} & \multicolumn{2}{c}{\textbf{Roundabout}} & \multicolumn{2}{c}{\textbf{T-Junction}} \\
\cmidrule(lr){2-3} \cmidrule(lr){4-5} \cmidrule(lr){6-7}
\textbf{Metric} & A & C & A & C & A & C \\
\midrule
Stalled (\%) & 26.89 & 39.56 & 36.36 & 24.79 & 67.45 & 32.54 \\
Off-road (\%) & 45.51 & 32.13 & 0.25 & 0.81 & 37.04 & 20.86 \\
Distance (m) & 36.09 & 38.75 & 30.45 & 47.68 & 16.20 & 23.73 \\
Reward & 92.39 & 50.68 & 56.56 & 98.57 & $-$47.95 & $-$4.26 \\
Velocity (km/h) & 4.41 & 4.77 & 3.87 & 6.05 & 2.14 & 3.17 \\
\bottomrule
\end{tabular}
\end{table}

\begin{figure}[t]
\centering
\includegraphics[width=0.8\textwidth]{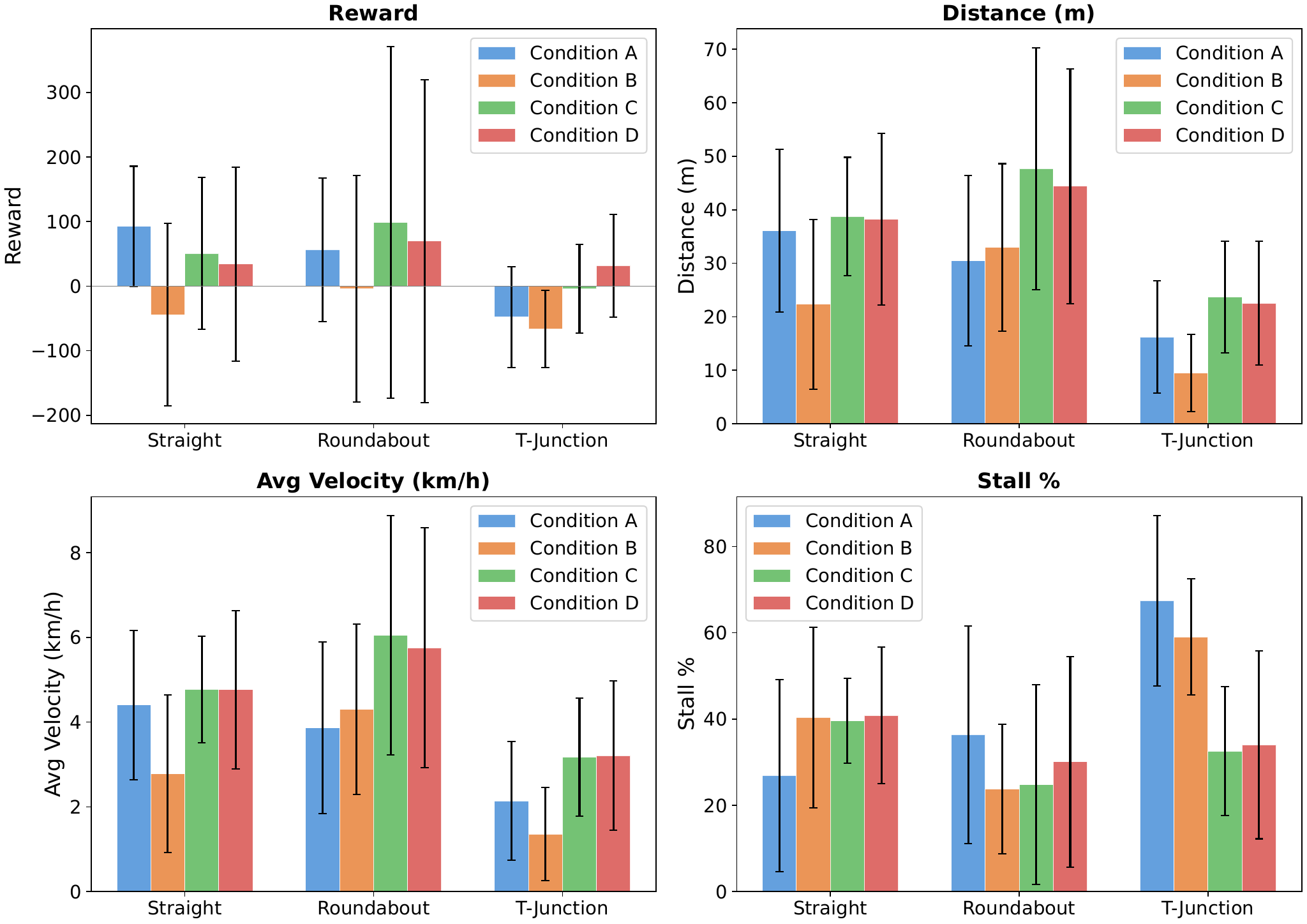}
\vspace*{-2ex}
\caption{Comparative performance across four ablation conditions. Each bar represents the mean value over 20 episodes, with error bars indicating standard deviation.}
\label{fig:four_condition_overview}
\end{figure}

To answer this RQ, Table~\ref{tab:rq2} compares Condition A (baseline) with Condition C (causal-trained policy + heuristic recovery). This represents the overall improvement of the hybrid system over the vanilla baseline.
As shown in Table~\ref{tab:rq2}, the hybrid system demonstrates clear performance gains in the roundabout and T-junction scenarios, with mixed results in the straight driving scenario:
\begin{enumerate}
    \item The roundabout scenario shows improvement in four of five metrics: stalled time decreases by 31.8\%, distance traveled increases by 56.6\%, cumulative reward rises by 74.3\%, and average velocity improves by 56.4\%.
    \item T-junction performance gains are even more pronounced, where all five metrics improve: reward increases by 91.1\%, stalled time falls by 51.8\%, off-road time drops by 43.7\%, and velocity rises by 48.0\%.
    \item The straight driving scenario yields mixed outcomes: while off-road time decreases by 29.4\%, stalled time increases by 47.1\% and reward declines by 45.1\%. Distance (+7.4\%) and velocity (+8.3\%) show marginal gains. We attribute this to the recovery module introducing stalled-state cycling in simpler environments where the baseline policy rarely needs intervention. %
\end{enumerate}
Figure~\ref{fig:four_condition_overview} presents a comparative visualization of all four ablation conditions across scenarios and performance metrics. %
 The bar chart confirms the quantitative gains reported above, with the most significant improvements of Condition~C over Condition~A observed in the T-junction (all five metrics improved) and roundabout (four of five metrics improved) scenarios, as detailed in the analysis above. 

\subsection{RQ3: External Recovery Contribution}\label{sec:rq3}

Table~\ref{tab:rq3} provides a quantitative analysis of the external assistance delivered by the heuristic recovery module under Condition C. 
Recovery interventions are observed across all evaluated scenarios. 
The roundabout scenario exhibits a notable pattern showing 9 of the 20 Condition~C episodes completed without requiring any recovery intervention. While a direct episode-level comparison with Condition~A is not straightforward (A has no recovery module to count interventions against), Condition~A's mean stalled time of 36.36\% indicates that the baseline policy regularly enters stalled states in this scenario. The fact that nearly half of Condition~C episodes avoid stalled states entirely suggests the causal-trained policy learned proactive avoidance behaviours.
By contrast, both the straight and T-junction scenarios necessitated recovery assistance in every episode, with the T-junction scenario demanding the highest intervention frequency (averaging 202.35 recovery actions per episode) owing to its elevated complexity.
\begin{table}[t]
\tablefontsize
\centering
\caption{Recovery Module Utilisation (Condition C).}
\label{tab:rq3}
\vspace*{-2ex}
\begin{tabular}{lccc}
\toprule
\textbf{Metric} & \textbf{Straight} & \textbf{Roundabout} & \textbf{T-Junction} \\
\midrule
Recoveries/episode & 7.40 & 5.15 & 8.70 \\
Successful recoveries & 3.75 & 0.35 & 1.65 \\
Success rate (\%) & 62.85 & 6.42 & 16.54 \\
Interventions (actions) & 109.10 & 136.20 & 202.35 \\
Avg recovery duration (steps) & 11.67 & 14.56 & 25.05 \\
Zero-recovery episodes & 0/20 & 9/20 & 0/20 \\
\bottomrule
\end{tabular}
\end{table}
Recovery success rates vary substantially across scenarios (Table~\ref{tab:rq3}). The straight scenario achieves 62.85\% success, while the roundabout (6.42\%) and T-junction (16.54\%) are notably lower. This disparity reflects the increased difficulty of reversing out of stalled states in geometrically complex environments: roundabouts involve curved road boundaries that limit viable reverse trajectories, while T-junctions present perpendicular road layouts where a simple reverse maneuver often leads to another stalled state or off-road position. However, even unsuccessful recovery attempts provide value as they generate exploratory movement that allows the PPO policy to resume forward driving from a different position. The high intervention counts (136.20 and 202.35 actions per episode) confirm that the recovery module remains actively engaged despite the low per-recovery success rate, and the overall performance gains for Conditions C and D in these scenarios (Section~\ref{sec:rq2}) demonstrate that the combined system still delivers substantial improvements.

The same results can be seen in Figure~\ref{fig:training_progress} and~\ref{fig:recovery_dynamics} where V2 (causal-guided PPO with recovery) achieves approximately 4.5× higher rewards than V1 (baseline without recovery) over 2000 training episodes. The policy improves navigation while maintaining steady recovery interventions ($\sim$2–3 per episode) with fluctuating success rates (40–80\%). The details are discussed in Section~\ref{sec:training_convergance} and Section~\ref{sec:Recovery_dynamics}.

\subsection{RQ4: Causal Training Effect on Policy Quality}
This RQ focuses on  our primary contribution that is showing whether causal-guided training enables the policy to develop better driving behaviors that extend beyond the capabilities of the recovery module alone. To address this we have analysed the results in multiple forms:

\begin{table}[b]
\tablefontsize
\centering
\caption{Performance in zero-recovery episodes (roundabout). V2 outperforms V1 without any recovery assistance.}
\label{tab:zero_recovery}
\vspace*{-2ex}
\begin{tabular}{lccc}
\toprule
\textbf{Metric} & \textbf{A (no recovery)} & \textbf{C (0 used)} & \textbf{Improvement} \\
\midrule
Distance (m) & 36.36 & 66.53 & +83.0\% \\
Reward & 94.43 & 337.51 & +257.4\% \\
Stalled (\%) & 30.56 & 5.11 & $-$83.3\% \\
\bottomrule
\end{tabular}
\end{table}

\begin{figure*}[t]
\centering
\includegraphics[width=\textwidth]{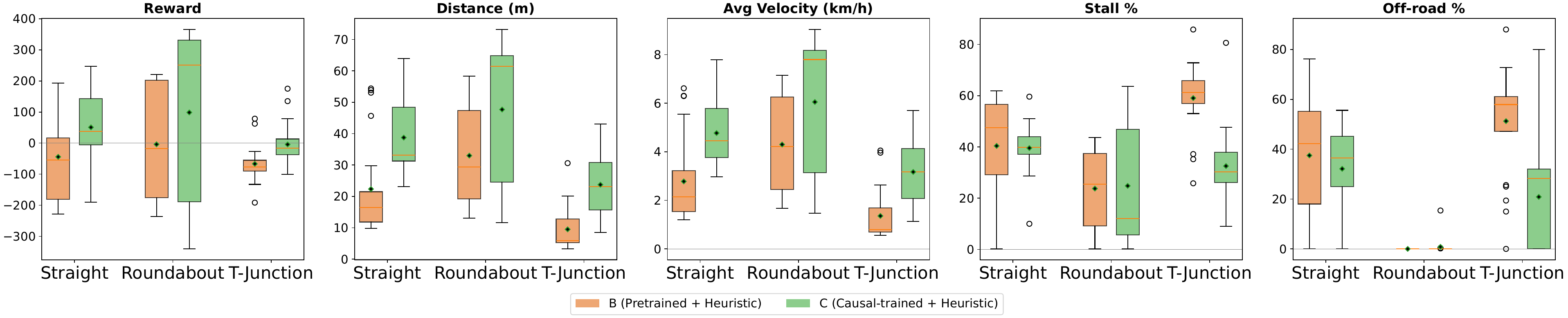}
\vspace*{-2ex}
\caption{C vs B: Effect of causal training (key comparison). Both conditions use identical heuristic recovery modules, differences reflect policy quality. Boxplots show per-episode distributions across 20 episodes. The causal-trained policy (C, green) consistently outperforms the pretrained baseline (B, orange) in reward, distance, and velocity across all scenarios.}
\label{fig:cvsb_boxplots}
\end{figure*}

\head{Zero-Recovery Episode Analysis (Roundabout)} 
In the roundabout scenario, 9 of 20 episodes completed without recovery intervention. Table~\ref{tab:zero_recovery} compares Condition A (baseline) against Condition C (causal-trained). The results demonstrate that the causal-trained policy traveled 83.0\% farther, earned 257.4\% higher reward, and spent 83.3\% less time stalled, demonstrating that causal-guided training enhances intrinsic policy capabilities independent of external interventions. To visualize the results, Figure~\ref{fig:cvsb_boxplots} presents the C vs B comparison, the key ablation that isolates the effect of causal training by holding the recovery module constant. Across all three scenarios, the causal-trained policy (C, green) shows higher medians and wider upward tails in reward, distance, and velocity, while exhibiting lower stalled percentages and off-road rates.
Interestingly, Condition B (pretrained policy with heuristic recovery) performs worse than Condition A (pretrained policy without recovery) across several metrics, particularly in the straight scenario (reward: -45.1\%, distance: -38.9\% which is confirmed by Mann-Whitney U tests with large negative effect sizes, Section~\ref{sec:stats}). This counterintuitive finding arises because the pretrained policy was never exposed to recovery interventions during training, thus cannot adapt when the recovery module assumes control. Following a recovery maneuver that repositions the vehicle, the pretrained policy reproduces the same actions that initially triggered the stalled state, resulting in a counterproductive stalled-recovery loop. This demonstrates why causal-guided training (Condition C) is essential. It enables the policy to cooperate with the recovery module, producing the substantial C vs B improvements shown in Figure~\ref{fig:cvsb_boxplots}.

\head{Stalled State Reduction} The results in Table~\ref{tab:rq2} show that the causal-trained policy (Condition C) consistently reduces stalled time relative to the baseline (Condition A) in complex scenarios, 31.8\% in the roundabout and 51.8\% at the T-junction. The straight driving scenario shows a counterintuitive 47.1\% increase in stalled time, likely due to recovery maneuvers contributing to stalled-state cycling in this simpler environment.%

\head{Off-Road Reduction} 
Table~\ref{tab:rq2} shows that off-road driving decreases in two of three scenarios: a 29.4\% reduction in straight driving and a 43.7\% reduction at the T-junction. This improvement during forward (non-recovery) driving indicates that the policy has learned effective navigation and lane-keeping behaviors.

\head{Velocity Maintenance} 
As shown in Table~\ref{tab:rq2}, the causal-trained policy achieves higher average velocity across all scenarios: +8.3\% in straight driving, +56.4\% in the roundabout, and +48.0\% at the T-junction. This velocity increase reflects improved forward progress and reduced time in low-velocity stalled states.

\begin{table}[b]
\tablefontsize
\centering
\caption{Full system (D) vs heuristic-only (B) across all scenarios. Both conditions have recovery modules, differences reflect policy quality and recovery strategy.}
\label{tab:d_vs_b}
\vspace*{-2ex}
\begin{tabular}{lcccccc}
\toprule
 & \multicolumn{2}{c}{\textbf{Straight}} & \multicolumn{2}{c}{\textbf{Roundabout}} & \multicolumn{2}{c}{\textbf{T-Junction}} \\
\cmidrule{2-7}
\textbf{Metric} & B & D & B & D & B & D \\
\midrule
Stalled (\%) & 40.39 & 40.84 & 23.75 & 30.07 & 59.04 & 34.00 \\
Off-road (\%) & 37.49 & 27.01 & 0.00 & 0.02 & 51.26 & 28.85 \\
Distance (m) & 22.34 & 38.20 & 32.98 & 44.38 & 9.49 & 22.53 \\
Reward & $-$44.33 & 34.06 & $-$3.89 & 69.92 & $-$66.67 & 31.69 \\
Velocity (km/h) & 2.78 & 4.77 & 4.30 & 5.75 & 1.36 & 3.21 \\
\bottomrule
\end{tabular}
\end{table}

\begin{figure*}[t]
\centering
\includegraphics[width=\textwidth]{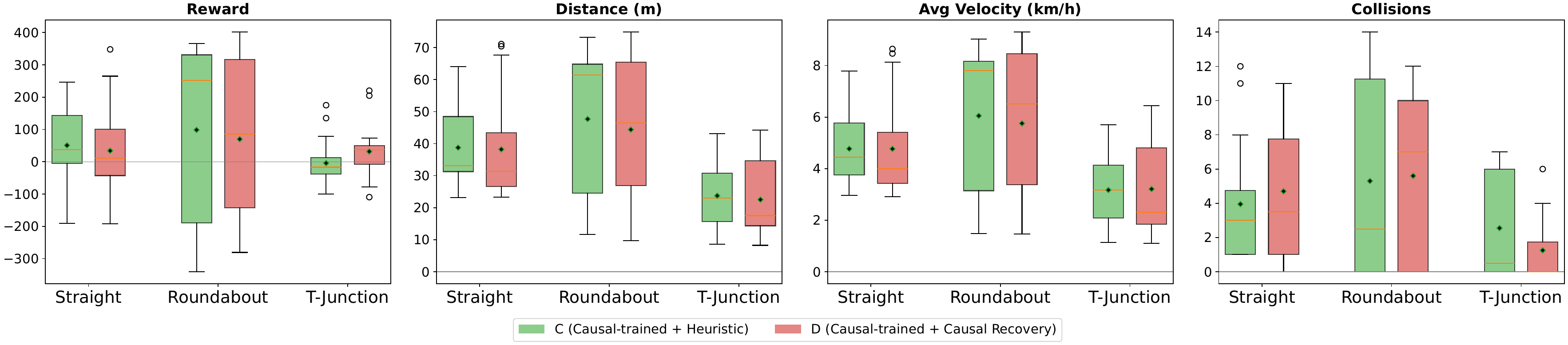}
\vspace*{-2ex}
\caption{D vs C: Effect of causal inference at runtime. Both conditions use causal-trained policy weights, D additionally uses the causal model for risk-ranked recovery action selection. Similar distributions across most metrics indicate successful knowledge transfer during training, with D showing modest improvements in collision reduction.}
\label{fig:dvsc_boxplots}
\end{figure*}

\head{Full System (D) vs Heuristic-Only (B) (Ablation)}
We compare Condition D (causal-trained policy with causal inference recovery) against Condition B (pretrained policy with heuristic recovery). Both use recovery mechanisms. Table~\ref{tab:d_vs_b} presents these comparative results. Since both conditions incorporate recovery modules, performance differences stem from two factors, policy quality from causal training and recovery strategy effectiveness. Condition D consistently outperforms B in straight driving showing 4 of 5 metrics improve (+176.8\% reward, +71.0\% distance, -28.0\% off-road time, +71.5\% velocity), in the roundabout the 3 of 5 metrics improve (+1897.4\% reward, +34.6\% distance, +33.8\% velocity),and at the T-junction, all 5 metrics improve (+147.5\% reward, -42.4\% stalled time, -43.7\% off-road time, +137.4\% distance, +136.3\% velocity). Figure~\ref{fig:dvsc_boxplots} examines whether keeping the causal model at inference time (Condition D) provides additional benefit beyond causal training with heuristic recovery alone (Condition C). The distributions are largely similar, with D showing slightly fewer collisions in some scenarios. This indicates that causal knowledge acquired during training transfers into the policy network weights.

\head{Recovery Quality: Causal vs Heuristic} 
Table~\ref{tab:recovery_quality} contrasts the recovery strategies employed in Conditions B (heuristic cycling) and D (causal model with risk-ranked actions). The causal model approach demonstrates superior efficiency. It reduces recovery attempts by 28-39\% in straight driving and T-junction scenarios, attains higher success rates in two scenarios (straight: 46.89\% versus 30.27\% and roundabout: 5.65\% versus 1.67\%), and completes recoveries 8--25\% faster across all scenarios. These results indicate that risk-ranked action selection guided by the causal model outperforms blind heuristic cycling.

\begin{table}[t]
\tablefontsize
\centering
\caption{Recovery module comparison: heuristic cycling (B) vs causal risk-ranked (D).}
\label{tab:recovery_quality}
\vspace*{-2ex}
\begin{tabular}{lcccccc}
\toprule
 & \multicolumn{2}{c}{\textbf{Straight}} & \multicolumn{2}{c}{\textbf{Roundabout}} & \multicolumn{2}{c}{\textbf{T-Junction}} \\
\cmidrule{2-7}
\textbf{Metric} & B & D & B & D & B & D \\
\midrule
Recoveries/ep & 10.25 & 7.30 & 5.05 & 5.30 & 12.70 & 7.75 \\
Success rate (\%) & 30.27 & 46.89 & 1.67 & 5.65 & 31.44 & 26.27 \\
Actions/ep & 232.65 & 131.30 & 140.75 & 142.80 & 293.80 & 164.90 \\
Avg duration (steps) & 21.07 & 15.77 & 20.52 & 16.32 & 20.72 & 19.15 \\
\bottomrule
\end{tabular}
\end{table}

\head{Zero-Recovery Episodes in D vs B (Roundabout)} 
In the roundabout scenario, both conditions B and D had episodes completing without recovery intervention (B: 6/20; D: 7/20). Analyzing these zero-recovery episodes isolates pure policy quality. Without recovery assistance, the causal-trained policy (condition D) shows superior performance, +24.8\% distance, +63.1\% reward, -17.2\% stalled time. This confirms that causal training enhanced intrinsic policy capabilities as shown in Table~\ref{tab:rq4}.

\begin{table}[b]
\tablefontsize
\centering
\caption{Zero-recovery episode comparison (roundabout, B vs D). Differences are purely from policy weights.}
\label{tab:zero_recovery_bd}
\vspace*{-2ex}
\begin{tabular}{lccc}
\toprule
\textbf{Metric} & \textbf{B (0 recovery)} & \textbf{D (0 recovery)} & \textbf{Improvement} \\
\midrule
Distance (m) & 47.09 & 58.77 & +24.8\% \\
Reward & 209.57 & 341.80 & +63.1\% \\
Stalled (\%) & 7.73 & 6.40 & $-$17.2\% \\
\bottomrule
\end{tabular}
\label{tab:rq4}
\end{table}

\subsection{RQ5: What are the collision trade-offs of recovery-enabled driving?}
Table~\ref{tab:rq5} presents the collision trade-off analysis across conditions and scenarios. The C vs A comparison (hybrid with recovery module versus baseline) shows that adding recovery substantially increases collisions (155-607\%), and the net reward trade-off is negative in two of three scenarios. However, the D vs B comparison (both have recovery, isolating causal training and inference) shows consistently positive net reward across all three scenarios (+58.39, +68.81, +92.36), accompanied by only moderate collision increases (4.7-31.6\%). This shows that when recovery capability is held constant, the causal system achieves net-positive performance gains. The increase in collisions in the C versus A comparison originates primarily from the recovery maneuvers themselves, rather than deficiencies in the causal policy.

\begin{table}[t]
\tablefontsize
\centering
\caption{Collision analysis and net reward trade-off. Recovery increases collisions but may yield net positive reward.}
\label{tab:rq5}
\vspace*{-2ex}
\begin{tabular}{lcccccc}
\toprule
 & \multicolumn{3}{c}{\textbf{C vs A}} & \multicolumn{3}{c}{\textbf{D vs B}} \\
\cmidrule(lr){2-4} \cmidrule(lr){5-7}
\textbf{Scenario} & 
\textbf{Collision $\Delta$} & \textbf{Net Reward} & \textbf{Verdict} &
\textbf{Collision $\Delta$} & \textbf{Net Reward} & \textbf{Verdict} \\
\midrule
Straight & +364.7\% & $-$103.70 & Negative & +27.0\% & +58.39 & Positive \\
Roundabout & +606.7\% & $-$48.99 & Negative & +4.7\% & +68.81 & Positive \\
T-Junction & +155.0\% & +12.69 & Positive & +31.6\% & +92.36 & Positive \\
\bottomrule
\end{tabular}
\end{table}

\begin{figure*}[b]
\centering
\includegraphics[width=.8\textwidth]{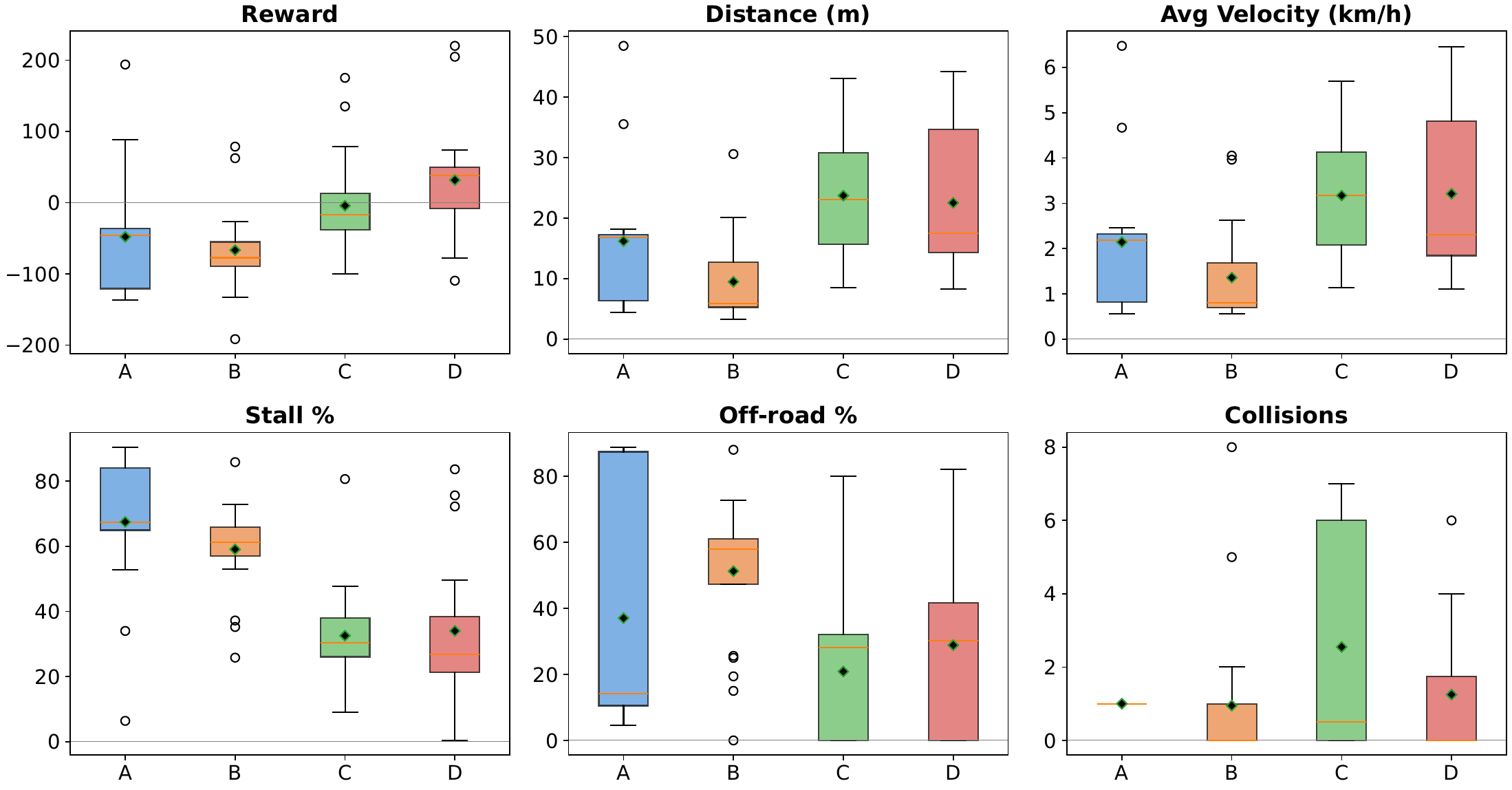}
\vspace*{-2ex}
\caption{T-junction: all four conditions (strongest scenario). Per-episode boxplots show progressive improvement from A to D. Conditions C and D achieve dramatically higher reward, distance, and velocity while reducing stalled and off-road percentages by approximately 50\% compared to A and B. Collision counts increase with recovery but are offset by forward-progress gains.}
\label{fig:merge_deep_dive}
\end{figure*}

Figure~\ref{fig:merge_deep_dive} offers an in-depth analysis of the T-junction scenario, which represents the most demanding test environment and demonstrates the most significant performance gains from the causal system. The boxplot distributions across all four experimental conditions reveal the incremental advantage of each system component: incorporating recovery capability (transition A$\rightarrow$B) yields moderate improvements, implementing causal training (transition B$\rightarrow$C) produces substantial enhancements across all performance metrics, and applying causal inference (transition C$\rightarrow$D) delivers additional refinement characterized by decreased collision variability.

\subsection{Statistical Analysis}\label{sec:stats}
To evaluate reliability, we employ the Mann-Whitney U test with sample sizes of 20 episodes per experimental condition, accompanied by Cohen's $d$ effect size measurements where $|d| < 0.2$ indicates negligible effects, $0.2$--$0.5$ represents small effects, $0.5$--$0.8$ denotes medium effects, and $\geq 0.8$ signifies large effects. Cohen's $d$ values for all four pairwise comparisons across scenarios are reported in Table~\ref{tab:stat_all}, with statistical significance marked by asterisks:
\begin{itemize}
    \item \textbf{C vs B (key comparison).} This ablation holds recovery constant and isolates causal training. The T-junction yields significance across all five metrics with large effects ($d = 0.97$--$1.86$), while distance and velocity reach significance in all three scenarios. Critically, collision counts show no significant difference (all $p > 0.15$), confirming that causal training improves performance without additional collision risk.

\item \textbf{B vs A.} Adding recovery alone \textit{degrades} straight-driving performance (reward: $d = -1.14^{**}$, distance: $d = -0.89^{**}$), proving that the recovery module's value is realised only when paired with causal-trained policies.

\item \textbf{D vs C.} Nearly all comparisons are non-significant with negligible effect sizes, supporting a knowledge distillation interpretation: causal guidance transfers into policy weights during training, eliminating the need for inference-time causal computation.

\item \textbf{D vs A.} The full system achieves significant improvements at the T-junction (reward: $d = 1.01^{***}$, stalled: $d = -1.61^{***}$) and roundabout (distance: $d = 0.73^{*}$, velocity: $d = 0.77^{*}$). Collision counts increase significantly ($p < 0.05$) but with a negligible effect size at the T-junction ($d = 0.18$), where performance gains are largest.

\end{itemize}

The C vs B comparison achieves the highest number of significant results (10 out of 15 metric-scenario combinations), with the T-junction producing significance across all five core metrics. This provides strong statistical evidence that causal-guided training genuinely improves policy quality beyond what the recovery module alone provides. 

\begin{table*}[tb]
\tablefontsize
\centering
\caption{Cohen's $d$ effect sizes for all ablation comparisons across scenarios and metrics. Mann-Whitney U test, $n=20$. Positive $d$ indicates improvement for the first-named condition. $*$: $p<0.05$, $**$: $p<0.01$, $***$: $p<0.001$.}
\label{tab:stat_all}
\vspace*{-2ex}
\begin{tabular}{lcccccccccccc}
\toprule
 & \multicolumn{3}{c}{\textbf{B vs A}} & \multicolumn{3}{c}{\textbf{C vs B (key)}} & \multicolumn{3}{c}{\textbf{D vs C}} & \multicolumn{3}{c}{\textbf{D vs A}} \\
\cmidrule{2-13}
\textbf{Metric} & S & R & T & S & R & T & S & R & T & S & R & T \\
\midrule
Reward & $-$1.14$^{**}$ & $-$0.41 & $-$0.27 & 0.73$^{*}$ & 0.45 & 0.97$^{**}$ & $-$0.12 & $-$0.11 & 0.48$^{*}$ & $-$0.47 & 0.07 & 1.01$^{***}$ \\
Distance & $-$0.89$^{**}$ & 0.16 & $-$0.75$^{*}$ & 1.20$^{***}$ & 0.76$^{*}$ & 1.59$^{***}$ & $-$0.04 & $-$0.15 & $-$0.11 & 0.14 & 0.73$^{*}$ & 0.57 \\
Velocity & $-$0.90$^{**}$ & 0.21 & $-$0.62$^{*}$ & 1.25$^{***}$ & 0.71$^{*}$ & 1.44$^{***}$ & $-$0.00 & $-$0.10 & 0.02 & 0.20 & 0.77$^{*}$ & 0.67 \\
Stalled & 0.62 & $-$0.61 & $-$0.50$^{**}$ & $-$0.05 & 0.05 & $-$1.86$^{***}$ & 0.10 & 0.22 & 0.08 & 0.72$^{*}$ & $-$0.25 & $-$1.61$^{***}$ \\
Off-road & $-$0.30 & $-$0.33 & 0.47 & $-$0.24 & 0.33$^{*}$ & $-$1.36$^{***}$ & $-$0.23 & $-$0.33 & 0.32 & $-$0.70$^{*}$ & $-$0.30 & $-$0.26 \\
\midrule
Sig. count & \multicolumn{3}{c}{0, 0, 1 / 5} & \multicolumn{3}{c}{3, 2, 5 / 5} & \multicolumn{3}{c}{0, 0, 1 / 5} & \multicolumn{3}{c}{1, 2, 2 / 5} \\
\bottomrule
\end{tabular}
\vspace*{-3ex}
\end{table*}

\subsection{Summary of Key Findings}

The 4-condition ablation study produces the following results. \textbf{Baseline limitations (RQ1)} shows Vanilla PPO without recovery demonstrates 26-67\% stalled time across scenarios, confirming the necessity for recovery mechanisms. \textbf{Hybrid system gains (RQ2)} confirms the causal-trained hybrid system achieves reward improvements of 74-91\% in complex scenarios (roundabout, T-junction) relative to the vanilla baseline.\textbf{Recovery utilisation (RQ3)} proves recovery module is employed regularly but not in all cases 9/20 roundabout episodes succeeded without recovery, indicating learned obstacle avoidance. \textbf{Learned behaviours (RQ4)} Zero-recovery episodes show +257\% reward and $-$83\% stalled time vs baseline, confirming policy enhancement independent of external intervention. The D vs B ablation validates that causal training improves policy quality +147-1897\% reward, +34-137\% distance, and +34--136\% velocity when controlling for recovery capability. Lastly, \textbf{Acceptable trade-offs (RQ5)} D vs B comparison produces net-positive reward in all scenarios despite modest collision increases, demonstrating that the trade-off is worthwhile.

\section{Limitations and Threats to Validity}

\head{Limitations}
This study has the following limitations: (1) the hybrid architecture relies on hand-engineered components for triggering, executing, and selecting recovery actions rather than learning them autonomously, (2) recovery maneuvers significantly increase collisions (155–607\%), with collision counts of 3.95–5.60 per episode that are too high for safe deployment, (3) causal training effectiveness varies by scenario, producing strong improvements at T-junctions but mixed or negative results in simpler straight driving conditions, (4) high episode-to-episode variance and limited sample size (n=20) reduce statistical power for detecting smaller effects, and  (5) experiments use only one CARLA map (Town07) with fixed conditions, leaving generalization untested.

\head{Internal Validity}
Threats to the internal validity include confounding between causal training and recovery exposure. Conditions C and D may reflect adaptation to recovery rather than superior driving. Non-independent episodes within CARLA sessions may introduce serial correlations. Another limitation is velocity threshold sensitivity, the <1.0 km/h stalled trigger inflates stalled percentages during recovery. A final threat stems from multiple comparisons across the 60 Mann--Whitney tests in Table~\ref{tab:stat_all} for which no correction was applied. The borderline single-star results ($p<0.05$) should therefore be interpreted with caution. The central C\,vs\,B T-junction findings reach the most stringent reporting level ($p<0.001$) with large effect sizes ($|d| = 1.36$--$1.86$), making them the least likely to be artifacts of multiple testing; a full family-wise correction is left to future analysis.

\head{External Validity}
All experiments were conducted in a single CARLA map (Town07) under fixed weather and lighting conditions with predetermined spawn points, which limits the generalizability of findings across different environments and scenarios. The sim-to-real gap presents an additional constraint that is CARLA does not fully model sensor noise, actuator dynamics, or unpredictable human behavior, and the collision increases observed under recovery conditions (155–607\% in C vs A) would necessitate additional safety mechanisms before real-world deployment. Furthermore, the framework was evaluated exclusively using PPO whether causal-guided training confers similar benefits to off-policy or model-based algorithms remains an open question for future investigation.

\head{Construct Validity}
We consider four threats to construct validity: First, cumulative reward serves as a hand-tuned proxy for driving quality; alternative penalty weightings could alter the relative ranking of conditions, particularly for recovery-enabled configurations. Second, the stalled-time metric can conflate genuinely immobilized states with legitimate low-speed driving behavior. Third, the binary recovery success criterion (velocity exceeding threshold) does not capture post-recovery positioning quality, potentially underestimating true recovery effectiveness. Fourth, the C vs A comparison used for RQ2 combines causal training and recovery presence into a single composite measure; ideally, component-level attribution should rely only on the C vs B ablation.

\section{Concluding Remarks}\label{sec:con}

\head{Conclusions}
This paper presents the Causal Recovery Reinforcement Learning (\method{}), a hybrid approach combining causal-guided reinforcement learning training with rule-based recovery to manage stalled states in autonomous vehicles. Using a comprehensive 4-condition ablation study in three CARLA driving scenarios (straight road, roundabout, T-junction), we systematically evaluate each component's impact and answer five key research questions.

Our findings show that standard PPO without recovery experiences 26–67\% stalled time across scenarios (RQ1), establishing the necessity for intervention mechanisms. The hybrid approach proves highly effective, delivering 74–91\% reward gains in complex scenarios (RQ2). Recovery assistance occurs 5–9 times per episode, yet the policy operates independently in 45\% of roundabout episodes (RQ3).

Our primary contribution is the causal-guided training methodology, which leverages a Bayesian causal model extracted from multi-agent driving interactions to inform RL policy development. By comparing Condition C versus B, where recovery remains constant, we see the compelling statistical evidence that causal training inherently enhances policy performance (RQ4). Mann-Whitney U tests reveal significant large-effect improvements in distance ($p < 0.001$, $d = 1.59$), velocity ($p < 0.001$, $d = 1.44$), stalled time ($p < 0.001$, $d = -1.86$), and off-road behavior ($p < 0.001$, $d = -1.36$) at the T-junction, with 10 of 15 metric scenario pairs achieving significance. Additionally, causal model-guided recovery surpasses heuristic cycling approaches, reducing recovery attempts by 28–39\% and recovery duration by 8–25\%. Though recovery maneuvers increase collisions, the substantial gains in forward progress produce net-positive rewards across all scenarios in the D versus B comparison (RQ5).

\head{Future Work}
Future work can address these limitations through several directions. Reverse actions can be integrated directly into the RL action space via curriculum learning to eliminate hand-engineered triggers. Another can be causal model collision predictions, which can be used to constrain recovery action selection within a Constrained MDP formulation. Adaptive duration, learned termination conditions, and recovery as a learned sub-policy can be implemented. Finally, experiments with larger sample sizes and robust statistical analysis should be conducted to improve confidence in results.

\printbibliography

 \end{document}